\documentclass{aa}
\usepackage{graphicx}
\usepackage{txfonts}
\usepackage{lipsum}
\usepackage{subcaption}         
\usepackage{lscape}             
\usepackage{placeins}           
\usepackage{hyperref}
\usepackage{makecell}
\usepackage[version=4]{mhchem}
\usepackage[dvipsnames]{xcolor}
\usepackage{dblfloatfix}
\usepackage{multirow}
\usepackage[utf8]{inputenc}
\usepackage{booktabs}
\usepackage{amssymb}
\usepackage{xspace}
\usepackage{array}
\newcommand{\micron}{ \textmu m}
\newcommand{\am}{I$_\textrm{a}$}
\newcommand{\cry}{I$_\textrm{c}$}
\newcommand{\hex}{I$_\textrm{h}$}

\begin{document}

\title{Implications of self-consistent \ce{H2O} ice optical constants on radiative transfer models of disks}

 \author{Z. L. Smith\inst{1}\corrauth{zsmith@strw.leidenuniv.nl}\and
          M. K. McClure\inst{1}\email{mcclure@strw.leidenuniv.nl} \and
          I. Kamp\inst{2}\email{kamp@astro.rug.nl} \and
          S. M. Cazaux\inst{3,1}\email{S.M.Cazaux@tudelft.nl}
          }

   \institute{Leiden Observatory, Leiden University, P.O. Box 9513, NL2300 RA Leiden, The Netherlands \and
    Kapteyn Astronomical Institute, University of Groningen, PO Box 800, 9700 AV Groningen, The Netherlands \and
    Faculty of Aerospace Engineering, Delft University of Technology, Delft, The Netherlands}

   \date{Received 27 March 2026; accepted 15 June 2026}
 
  \abstract
   {Interstellar water (\ce{H2O}) ice exhibits significantly varied profiles over its spectral features with temperature and thermal history. No previous radiative transfer model of a protoplanetary disk has fully accounted for these effects over the wavelength range from 3 - 200\micron\ simultaneously. A radiative transfer model with region-based distribution of ices and applicable ice optical constant composites (OCCs) are required to properly model the distribution and spectral signatures of water ice in disks.}
   {First, we build a series of thermal history- and temperature-dependent \ce{H2O} ice OCCs with the available literature data from 0.1 - 10,000\micron. We then self-consistently apply these within a radiative transfer model of a protoplanetary disk according to its thermal structure. Second, we assess the impact that appropriate \ce{H2O} optical constants have on the 3\micron\ absorption and 45 and 63\micron\ emission profiles in disks populated by ice with one of the two thermal history models.}
   {We combine multiple experiment measurements of the imaginary refractive indices, $k$, to cover from 0.1 - 10,000\micron\ and perform Kramers-Kronig (KK) integrations to derive updated real refractive indices, $n$. We construct \ce{H2O} ice OCCs for two thermal histories: first, direct deposit and measure at a fixed temperature ices (\textbf{\textit{DT}}) and deposit and cool down (\textbf{\textit{DC}}) ices, for five and eight distinct temperatures, respectively. We run four \texttt{RADMC-3D} radiative transfer models of edge-on class II disk, HH 48 NE. The four models each use different OCCs; the standard single temperature crystalline (model M0) and amorphous (model M0A) \ce{H2O} ice OCCs available from \texttt{OpTool} and our own multi-temperature \textbf{\textit{DT}} set of OCCs (model MDT) and \textbf{\textit{DC}} only set of OCCs (model MDC).}
   {We present the first radiative transfer models with temperature-dependent ice opacities over the full wavelength range relevant for \ce{H2O} ice. The 3\micron\ absorption feature shifts from 3.0\micron\ to 3.1\micron\ and shows a 3.2\micron\ shoulder when using the \textbf{\textit{DC}} OCCs, while the 45/63\micron\ emission features only appear when using the \textbf{\textit{DC}} OCCs.}
   {Using our new OCCs and temperature-specific opacity zoning within RT models together provides the first self-consistent RT profiles for the two most commonly used spectroscopic tracers of thermal processing of \ce{H2O} ice. We find that ices must have been thermally processed at temperatures higher than the local disk temperature in order to produce crystalline spectral profiles at 3, 45, or 63\micron\ in disks, suggesting local heating events or outward transport to colder regions, confirming the conclusions of previous radiative transfer models of edge-on classical T Tauri disks.}

    \titlerunning{Implications of self-consistent \ce{H2O} ice optical constants on radiative transfer models of disks}
    \authorrunning{Smith, Z. L. et al. 2026}

   \maketitle
   \nolinenumbers

\section{Introduction}
\label{sec:Intro}

Water (\ce{H2O}), an essential ingredient for life, is thought to be incorporated into forming planets and moons  as icy dust grains. \ce{H2O} ice is observed across all stellar formation stages, from molecular clouds \citep{boogert_ice_2011,mcclure_ice_2023,smith_cospatial_2025} to disks \citep{boogert_c2d_2008,mcclure_detections_2015,sturm_jwstmiri_2024,potapov_simple_2025}, comets \citep{fornasier_rosettas_2016} and the surfaces of our Solar System's icy moons \citep{villanueva_jwst_2023,bockelee-morvan_composition_2024,cartwright_jwst_2025}, through its vibrational absorption bands, including the 1–2\micron\ overtones, the 3\micron\ stretching mode, the 4.5\micron\ combination band, the 6\micron\ bending mode, the 11\micron\ libration mode, and the far-infrared (FIR) 45 and 63\micron\ emission ice lattice modes.
In planet-forming disks, ices may enhance the sticking properties of dust grains \citep{musiolik_contacts_2019}, and alter the disk's thermal structure due to its optical properties. 

The shapes of the 3, 45 and 63\micron\ bands reveal \ce{H2O} ice’s thermal history, as structural changes in the ice matrix alter the spectral profiles \citep{smith_molecular_1994,mastrapa_optical_2009}. The amorphous-to-crystalline phase transition produces characteristic changes: the 3\micron\ band narrows and shifts from 3.0 to 3.1\micron\ with the emergence of a 3.2\micron\ peak, while the FIR 45\micron\ band sharpens and the 63\micron\ feature appears only in crystalline ice \citep{smith_absorption_1989}. If ice crystallises and subsequently cools, it retains its crystalline structure but subsequent peak strengths and wavelength shifts that track the amount by which it has cooled \citep{smith_molecular_1994,mastrapa_optical_2009}. These spectral differences allow thermal histories to be inferred observationally.

\ce{H2O} ice has been detected via these features towards a number of Class II classical (K0 and later spectral type) T Tauri protoplanetary disks \citep{aikawa_akari_2012,terada_discovery_2012,mcclure_probing_2012,mcclure_detections_2015,sturm_jwstmiri_2024,potapov_simple_2025}, with a mixture of amorphous and crystalline profiles being observed. This split in observed \ce{H2O} ice phase is still not fully understood. For disks with crystalline profiles, it has been consistently determined that the crystalline ice must reside in the disk's cooler, outer regions, where ice would not natively crystallise \citep{mcclure_detections_2015,min_abundance_2016,kamp_diagnostic_2018}. This implies that additional dynamical processes within the disk must either generate these crystalline ices locally or transport them there. Such dynamical processes as viscous spreading or extreme heating phenomena, such as an accretion heating event or planetesimal collisions have been speculated to be the cause.

Accurate interpretation of observed disk ice features requires radiative transfer (RT) modelling. Grain size and shape distributions modify ice opacities \citep{mcclure_probing_2012, dartois_spectroscopic_2024}, and disk geometry affects the resulting spectra \citep{sturm_jwstmiri_2024}. RT models therefore depend on reliable \ce{H2O} optical constants across the full wavelength range (0.1–1,000\micron) needed to compute both stellar wavelength absorption/scattering and long wavelength thermal emission. No single laboratory experiment spans this range, so OCCs are constructed from multiple datasets.

Commonly used OCCs include the 266~K crystalline dataset of \citet{warren_optical_2008} and an amorphous composite distributed through \texttt{OpTool} \citep{dominik_optool_2021}, which merges 10~K measurements from \citet{hudgins_mid-_1993} (2.5 – 90\micron) with \citet{warren_optical_2008} outside that range. Although widely adopted, these fixed-temperature OCCs are often applied uniformly across disk models despite strong radial temperature gradients. This approach neglects temperature dependence and thermal history effects, resulting in ice band profiles that reflect inconsistent ice structures between commonly studied bands, e.g. the 3\micron\ versus 63\micron\ features. Ad hoc substitution of individual laboratory datasets into these OCCs is common \citep{cheng_comparison_2010,sturm_jwstmiri_2024, thang_evidence_2024,bergner_jwst_2024}, but lacks standardisation and limits the consistency across studies. \citet{min_abundance_2016} and \citet{kamp_diagnostic_2018} both began to address the thermal history issue in their studies, by including temperature appropriate ice opacities throughout their disk models in a locally self-consistent way for the FIR features. With the advent of JWST and future FIR missions on the horizon, \ce{H2O} ice optical constants and radiative transfer models that treat ice temperatures in a globally self-consistent way across the full 3-200\micron\ wavelength range are an essential tool to probe different thermal histories of disks.

Here, we present new \ce{H2O} optical constant composites (OCCs) that are globally self-consistent over the 3-200\micron\ range for two thermal histories: deposit-and-measure at fixed temperature and deposit-and-cool-down, spanning five and eight temperatures. We extend these OCCs to 0.1 – 1,000\micron\ in order to implement them in the \texttt{RADMC-3D} radiative transfer code and expand the ice zoning in the established RT model of HH 48 NE \citep{sturm_edge-protoplanetary_2023-1,sturm_jwstmiri_2024} to include corresponding ice temperature regions. By comparing the resulting 3, 45, and 63\micron\ spectral profiles with those of models employing the standard \citet{warren_optical_2008} crystalline and \citet{hudgins_mid-_1993} \texttt{OpTool} amorphous OCCs, we assess whether crystalline features should be detectable when distributed according to the intrinsic disk temperature gradient. If not, this would imply that additional dynamical processes, such as viscous spreading or transient heating events, are required to bring or produce thermally processed ice in the outer regions of disks, in order to explain observed crystalline ice signatures in classical T Tauri disks.

\section{Building thermal history- and temperature-dependent OCCs of pure \ce{H2O} ice}
\label{sec:OCs}

In this section, we construct OCCs for $n$ and $k$ of the complex refractive index, $m$ = $n$ + i$k$ from 0.1 - 10,000\micron\ for pure \ce{H2O}. Two sets of OCCs are built, split by the thermal history of the experimentally grown ices. 

The first set of pure \ce{H2O} ice OCCs combine optical constant measurements of the deposit and measure at fixed temperatures thermal history for five temperatures and is subsequently denoted as the \textbf{\textit{DT}} set from herein. The low temperature measurements (30~K) will emulate an amorphous ice layer formed atop a dust grain in the early cloud stages of star formation or to the icy dust towards the outer regions of a disk. The higher temperature measurements (75~K and above) more closely emulate ice formation occurring the natively warmer inner regions of a disk. The second set of OCCs combines optical constant measurements of the deposit and cool down thermal history for eight temperatures and is subsequently denoted as the \textbf{\textit{DC}} set from herein. All \textbf{\textit{DC}} experimental measurements across the full wavelength range deposit the ice at 150~K and are then cooled, therefore they are all crystalline in ice phase. These measurements allow us to emulate interstellar ices that have undergone thermal processing in the inner disk regions and have been transported to the outer, cold regions. 

The laboratory experimental data used to build our OCCs are all performed under slightly different conditions, which are discussed in the following sections, but all produce transmission spectra and then derive the imaginary component of the refractive index, $k$, values using the transmission spectra. In our OCCs, we collate $k$ values across different wavelength ranges for each temperature and thermal history combination and stitch them together. These datasets are shown in Table \ref{tab:combined_optical_constants}. In most cases, the $k$ values match across experiments but where discrepancies exist between experimental data, we perform interpolation to ensure a smooth transition from one dataset to another. The space in which the interpolation is chosen to be performed in aims to best match the overall profile of the already existing data. The instances where interpolation is used are outlined in the following sections. This same approach was employed in \citet{warren_optical_1984} when building the widely used \ce{H2O} at 266~K composite. Once we have the $k$ values in terms of wave number, $\nu$, we calculate the real component of the refractive index, $n$, using the the Kramers-Kronig (KK) relation:

\begin{equation}
\label{eq:KK}
    n(\nu) = n_{632\rm{nm}} + \frac{2}{\pi} \mathcal{P} \int_{v_1}^{v_2} \frac{\nu' k(\nu')}{\nu'^2 - \nu^2} d\nu
\end{equation}

where $n_{632\rm{nm}}$ is the real refractive index at the reference wavelength of 632~nm. For our \textbf{\textit{DT}} OCCs we use the directly measured $n_{632\rm{nm}}$ from \citet{he_refractive_2022}, which was measured to vary with temperature. However, for the \textbf{\textit{DC}} OCCs, no temperature-dependent measurements in this wavelength range have been taken, therefore, we use $n_{632\rm{nm}}$ = 1.32 from \citet{hale_optical_1973}. $\nu'$ is the wave number used to calculate the integral against the wave number being calculated for, $\nu$. The Cauchy Principle, $\mathcal{P}$, is used to overcome the singularity, $\nu' = \nu$. We chose to perform KK calculations using the $k$ values across the total wavelength range of the OCCs rather than for each dataset individually as would have been done in their original studies, as this will provide the most accurate $n$ values. To derive usable $n$ values, we perform the KK integral numerically using Simpson integration over a grid of 100,000 evenly spaced wave numbers in log($\nu$) and then we downsample these $n$ values back to the native frequency grid of each datasets experiments. \citet{warren_optical_1984} state that solving Equation \ref{eq:KK} with such fine gridding and singularity that is of type 0/0 means there is a minimal impact on the final KK integral if one performs an analytical integration at the singularity to account for the missing integral area. The KK integral runs from zero to infinity, meaning that the wider range of wavenumbers you have data for, the more accurate $n$ values can be derived. By creating these self-consistent OCCs, we are able to improve on the derivation of the $n$ values for each experimental dataset in isolation. An additional impact of the integral running from zero to infinity is that the ends of the calculated wavelength range suffer from truncation errors. Therefore, an estimate of $n$ values outside of our desired composite frequency range are required to ensure reliable $n$ values are derived for our end wavelengths, 0.1 and 10,000\micron. How we approach the truncation error is described in Section \ref{sec:common_data}. \\

\begin{table}[h!]
\centering
\caption{Laboratory datasets used to construct the fixed temperature thermal history (\textbf{\textit{DT}}) and cool down thermal history (\textbf{\textit{DC}}) \ce{H2O} OCCs.}
\label{tab:combined_optical_constants}
\setlength{\tabcolsep}{2.5pt}
\begin{tabular}{l l l}
\toprule
Wavelength & \textbf{\textit{DT}} composite & \textbf{\textit{DC}} composite\\
range (\textmu m) & experimental dataset & experimental dataset\\
\toprule

0.1 - 1.2 
& \multicolumn{2}{c}{\citet{warren_optical_2008}} \\

\midrule

1.2 - 2.64 
& \citet{mastrapa_optical_2009}
& \multirow{7}{*}{\citet{mastrapa_optical_2009}} \\

\cmidrule(lr){2-2} 

2.64 - 2.82
& Log-log interpolation
& \\

\cmidrule(lr){2-2} 

2.82 - 15 

& \multirow{4}{*}{\makecell[l]{\citet{rocha_water_2024}; \\ \citet{gerakines_ultraviolet_1996} \\ 160~K;\\ \citet{oberg_effects_2007} \\ $<$160~K}}

&  \\

& & \\
& & \\
& & \\
& & \\

\cmidrule(lr){3-3}

15 - 18
& 
& \multirow{2}{*}{Log-log interpolation} \\

\cmidrule(lr){2-2} 

18 - 20 
& Log-log interpolation
& \\

\midrule

20 - 74.5 
& \makecell[l]{\citet{smith_molecular_1994} \\ $<$75~K; \\ \citet{curtis_measurement_2005} \\ $>$75~K}
& \citet{smith_molecular_1994} \\

\midrule

74.5 - 94 
& Linear interpolation
& Log-log interpolation \\

\midrule

94 - 10,000 
& \multicolumn{2}{c}{\citet{reinert_absorption_2015}} \\

\bottomrule
\end{tabular}
\tablefoot{All datasets except \citet{warren_optical_2008} contain temperature-dependent measurements. The interpolation method listed was chosen to best match the profile of the experimental data. Where references are listed in both \textbf{\textit{DT}} and \textbf{\textit{DC}} series, there are multiple thermal histories in that study, except \citet{warren_optical_2008}, which is \textbf{\textit{DT}} data and \citet{reinert_absorption_2015}, which is \textbf{\textit{DC}} data.}

\end{table}

Table \ref{tab:Temp_Phase} shows the temperatures available in each thermal histories' composite set and the according ice phase for each measurement. The number of temperatures we can build OCCs for is limited by the available spectroscopic data from laboratory experiments across the required wavelength range for RT modelling. When creating each set of \ce{H2O} ice OCCs, we aimed to match the temperatures across the full wavelength range and between thermal histories as closely as possible. These matched temperatures for the \textbf{\textit{DT}} and \textbf{\textit{DC}} OCCs can have a 5~K difference but this is deemed a reasonable match within the temperature range sampled. These matching temperature subsets are denoted by double-barred boxes in Table \ref{tab:Temp_Phase}. \\

\begin{table}[h]
    \centering
    \caption{The temperature, ice phase and thermal history for every each \ce{H2O} OCC we compose.}
    \label{tab:OCCs_Phase_TH}
    \begin{tabular}{|c|c|c|}
        \hline
        \makecell{Temperature \\ (K)} & Phase & \makecell{Thermal \\ History} \\
        \hline
        20 & \cry & \textbf{\textit{DC}} \\
        \hline
        \hline
        30 & \am & \textbf{\textit{DT}} \\
        30 & \cry & \textbf{\textit{DC}} \\
        \hline
        \hline
        40 & \cry & \textbf{\textit{DC}} \\
        \hline
        50 & \cry & \textbf{\textit{DC}} \\
        \hline 
        \hline
        75 & \am & \textbf{\textit{DT}} \\
        80 & \cry & \textbf{\textit{DC}} \\
        \hline
        \hline
        105 & \am & \textbf{\textit{DT}} \\
        110 & \cry & \textbf{\textit{DC}} \\
        \hline
        \hline
        135 & \cry & \textbf{\textit{DT}} \\
        140 & \cry & \textbf{\textit{DC}} \\
        \hline
        \hline
        150 & \cry & \textbf{\textit{DC}} \\
        160 & \cry & \textbf{\textit{DT}} \\
        \hline
    \end{tabular}
    \tablefoot{Temperatures have grouped into closest matches between the two measurement techniques, \textbf{\textit{DT}} and \textbf{\textit{DC}}. \am\ = amorphous phase and \cry\ = crystalline phase.}
    \label{tab:Temp_Phase}
\end{table}

In the following sections, for each set of thermal-history and temperature-dependent OCCs, we describe the details of the experimental setup, covered wavelength ranges and any required interpolation between datasets used to produce each OCCs final $k$ values. We also note wavelength regions where potential mismatches in either thermal history, temperature or ice phase with the target thermal history and temperature for a composite exist and our chosen approach to ensure continuity. We describe the experimental datasets first followed by the interpolation between them, so the reader has the full experimental data context. Figures showing the interpolation we perform can be found in Appendices \ref{sec:AppA_DT} and \ref{sec:AppA_DC}.

\subsection{Common \textbf{\textit{DT}} \& \textit{\textbf{DC}} experimental data}
\label{sec:common_data}
Table \ref{tab:combined_optical_constants} shows that for both set of thermal history-dependent OCCs use the $k$ values from the same datasets as each other in the 0.1 - 1.2\micron\ and 94 - 10,000\micron\ wavelength ranges. This is due to very limited experimental data at these wavelengths. We describe the experimental data used and the $n$ values we calculate in the following sections.

\subsubsection{0.1 - 1.2\micron: \citet{warren_optical_2008}}
\label{sec:0.1_1.2}

The \citet{warren_optical_2008} composite was derived for pure \ce{H2O} ice in its hexagonal structure (\hex) at 266~K, with a focus on solar system and terrestrial ices measurements. From 0.04 - 0.16\micron, they use \citet{seki_optical_1981}'s experimental data of crystalline \ce{H2O} ice at 80~K, which remains unchanged from their original composite \citep{warren_optical_1984}. From 0.16 - 1.2\micron, they combine a complex series of experimental data and interpolation, so we refer the reader to the original \citet{warren_optical_2008} paper for the details to keep this paper within a reasonable scope. They address the clear mismatch of temperature in the 0.04 - 0.16\micron\ regime (80~K to the target 266~K) in \citet{warren_optical_1984} where they note that experiments have shown that the $k$ values appear to show very little variation with temperature or ice phase. This lack of temperature or ice phase dependence and a lack of more recent data in this wavelength regime, we use the \citet{warren_optical_2008} composite $k$ values for every \textbf{\textit{DT}} and \textbf{\textit{DC}} composite. Their composite extends far beyond 1.2\micron\ but only use to 1.2\micron\ as this is the wavelength from which much better suited data from \citet{mastrapa_optical_2009} is available. Finally, to overcome the truncation issue for the KK integral, we use the \citet{warren_optical_2008} data down to 0.04\micron\ when calculating our $n$ values and then cut our composite's $n$ values at 0.1\micron. 

\begin{figure*}[t]
   \centering
   \includegraphics[width=0.85\textwidth]{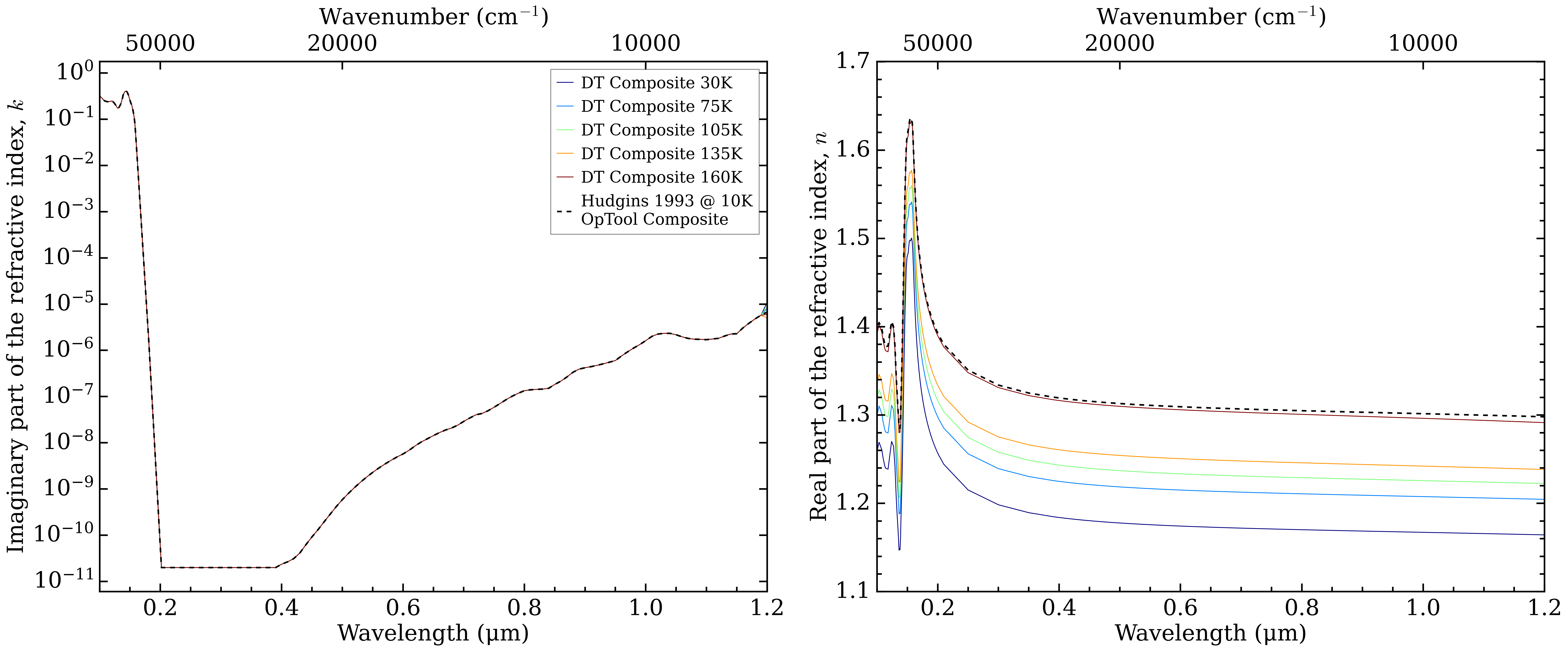}
              \caption{UV - NIR zoom in of the \textbf{\textit{DT}} OCCs at each temperature shown together with the \citet{hudgins_mid-_1993} \texttt{OpTool} composite, which in this region uses the \citet{warren_optical_2008} at 266~K data. 
              }
         \label{fig:DT_0.1_1.2}
\end{figure*}

\begin{figure*}[t]
   \centering
   \includegraphics[width=0.85\textwidth]{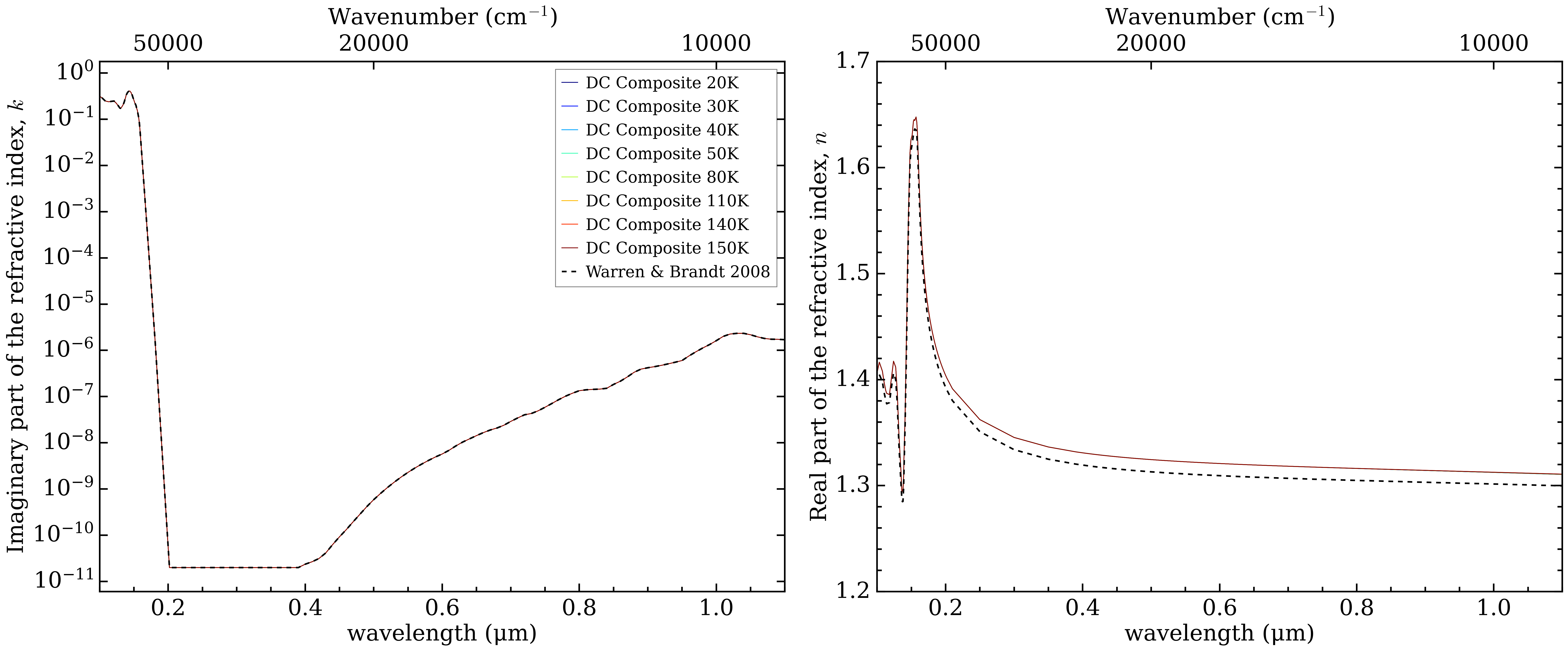}
              \caption{UV - NIR zoom in of the \textbf{\textit{DC}} OCCs at each temperature shown together with the \citet{warren_optical_2008} at 266~K data.
                      }
         \label{fig:DC_0.1_1.2}
\end{figure*}

Figures \ref{fig:DT_0.1_1.2} \& \ref{fig:DC_0.1_1.2} show the $k$ (left panels) and $n$ (right panels) values show the \textbf{\textit{DT}} and \textbf{\textit{DC}} derived values, respectively, for every temperature composite in their respective series. The $k$ values are identical for all OCCs but Figure \ref{fig:DT_0.1_1.2} demonstrates the impact of the temperature dependence of the $n_{632\rm{nm}}$ values as measured by \citet{he_refractive_2022} on each \textbf{\textit{DT}} composite, with Figure \ref{fig:DC_0.1_1.2} showing the \textbf{\textit{DC}} OCCs are fixed in place at $n_{632\rm{nm}}$ = 1.32 from \citet{hale_optical_1973}. \citet{he_refractive_2022}'s experiments measured $n$ values every 10~K from 30~K up to 160~K. Each temperature measurement was taken as the \ce{H2O} was deposited onto the aluminium mirror when it was at the fixed deposition temperature. 

\subsubsection{94 - 10,000\micron: \citet{reinert_absorption_2015}}
\label{sec:94_10000}

\citet{reinert_absorption_2015} built a temperature-dependent far-infrared analytical model of hexagonal crystalline ice \ce{H2O} (\hex) for wavelengths of 94\micron\ and above. Their experiments placed distilled liquid \ce{H2O} between two windows of polyethylene, \ce{(C2H4)_n} inside of a continuous flow helium cryostat and cooling the liquid \ce{H2O} to 250~K resulting in an \ce{H2O} \hex\ ice. They then cooled and measured their ice sample at six different temperatures; 250~K, 200~K, 150~K, 100~K, 50~K and 10~K. Spectroscopic measurements at each temperature were taken across three distinct wavelength ranges (80 - 192\micron, 149-385\micron, 225 - 625\micron). Each experimental temperature had a slightly different wavelength range measured within the three ranges, therefore the model is built only using the data where $\lambda$ > 94\micron\ to keep consistency at the shorter wavelengths. 

This dataset does match the thermal history of the \textbf{\textit{DC}} set, making it the ideal dataset to use. It does not match the thermal history of the \textbf{\textit{DT}} set, however, we chose to use their analytical model over other available data to ensure temperature-dependence in our OCCs. This highlights another significant missing temperature dependent measurement for the fixed temperature thermal history for \ce{H2O} ice. We use the \citet{reinert_absorption_2015} analytical models from 94\micron\ in every composite and cut off the shorter wavelength experimental data to ensure consistent temperature dependence across each set of OCCs. Additionally, we are able to generate $k$ values beyond the required 1,000\micron\ with the analytical models, therefore we do so out to 10,000\micron\ to overcome the truncation issue of the KK integral. 

\begin{figure*}[t]
   \centering
   \includegraphics[width=0.85\textwidth]{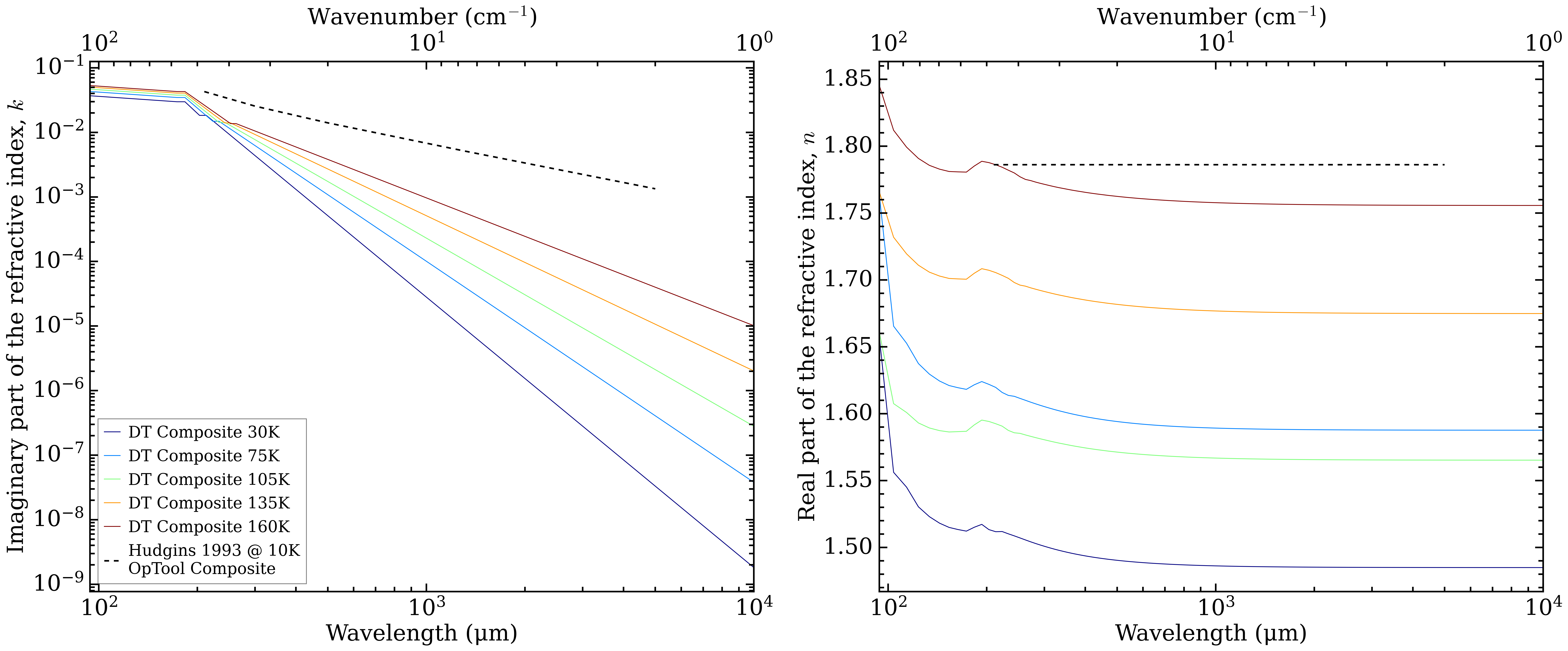}
      \caption{Beyond FIR zoom in of 94 - 10,000\micron\ of the \textbf{\textit{DT}} OCCs at each temperature shown together with the \citet{hudgins_mid-_1993} \texttt{OpTool} composite, which in this region uses the \citet{warren_optical_2008} at 266~K data. All our OCCs use the analytical model of \citet{reinert_absorption_2015}, deriving values at each composite temperature. 
              }
         \label{fig:DT_94_10000}
\end{figure*}

\begin{figure*}[t]
   \centering
   \includegraphics[width=0.85\textwidth]{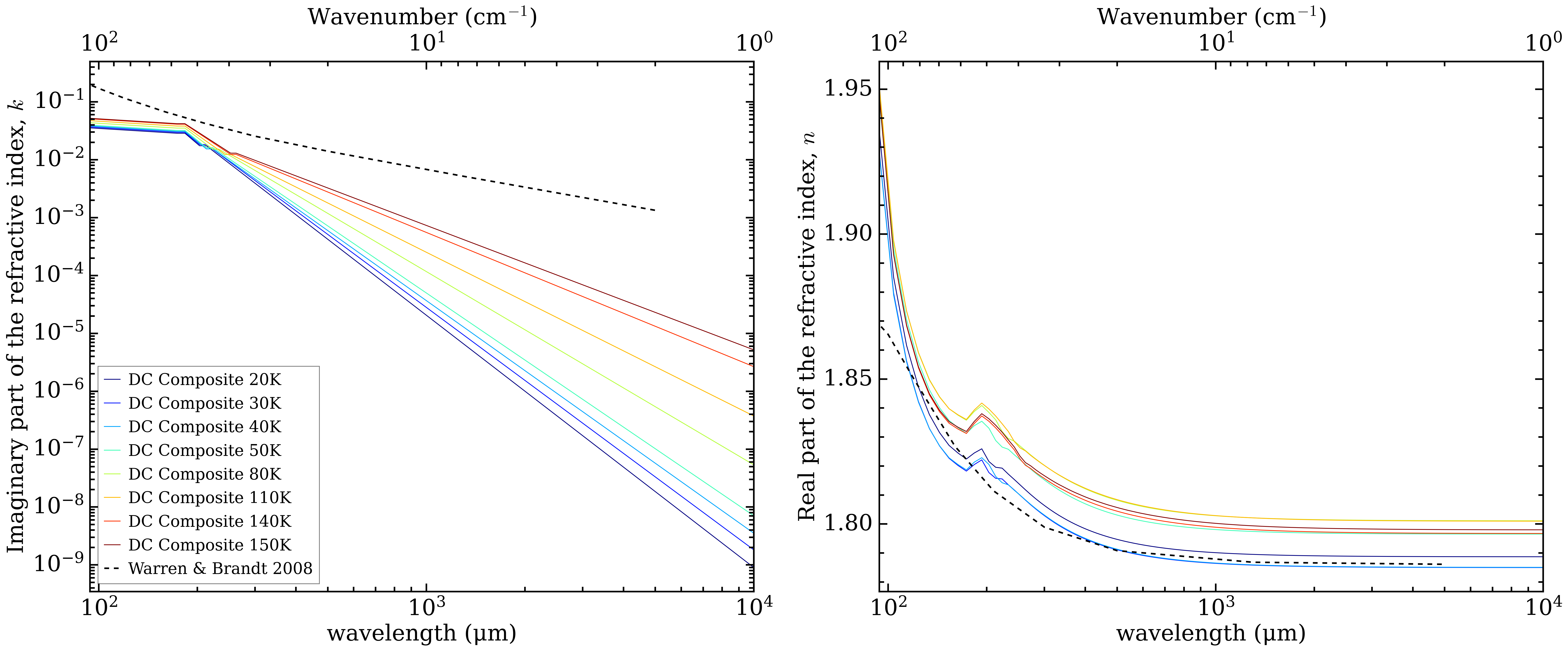}
      \caption{Beyond FIR zoom in of 94 - 10,000\micron\ of the \textbf{\textit{DC}} OCCs at each temperature shown together with the \citet{warren_optical_2008} at 266~K data. All our OCCs use the analytical model of \citet{reinert_absorption_2015}, deriving values at each composite temperature. 
              }
         \label{fig:DC_94_10000}
\end{figure*}

Figures \ref{fig:DT_94_10000} \& \ref{fig:DC_94_10000} show the $k$ (left panels) and $n$ (right panels) values show the \textbf{\textit{DT}} and \textbf{\textit{DC}} derived values, respectively, for every temperature composite in their respective series. The $>$94\micron\ values show much steeper slopes that the \citet{warren_optical_2008}, as shown in \citet{reinert_absorption_2015,kim_constraining_2019}. Due to two different datasets being used in the 20-74.4\micron\ range for the \textbf{\textit{DT}} OCCs, we see in Figure \ref{fig:DT_94_10000} that the temperature-dependent trend does not hold in the $n$ values beyond 94\micron. Until new self-consistent experimental data of the FIR region across our temperature range are measured, we are unable to address this.  

\subsection{Deposition temperature (\textbf{\textit{DT}}) measurement pure \ce{H2O} OCCs}
\label{sec:DT_exps}

\subsubsection{1.2 - 2.64\micron: \citet{mastrapa_optical_2009}}
\label{sec:DT_1.2_2.64}
We use the \citet{mastrapa_optical_2009} deposit and measure at a fixed temperature experimental data from 1.2 - 2.64\micron\ to include the \ce{H2O} overtone transitions in our \textbf{\textit{DT}} OCCs.
For these data, \citet{mastrapa_optical_2009} deposited water vapour at a series of temperatures (15, 25, 40, 50, 60, 80, 100, 120~K) onto either a potassium bromide (KBr), zinc selenide (ZnSe) or cesium iodide (CsI) substrate and immediately took the spectroscopic measurements. This means that for our 30~K and 75~K \textbf{\textit{DT}} OCCs, there is a 5~K mismatch at these wavelengths. Figure \ref{fig:DT_1.2_20} shows the temperature dependence of the \ce{H2O} overtone transitions in our OCCs.

\subsubsection{2.82 - 20\micron: \citet{gerakines_ultraviolet_1996,oberg_effects_2007,rocha_water_2024}}
\label{sec:DT_2.82_20}
\begin{figure*}[t]
   \centering

   \includegraphics[width=0.85\textwidth]{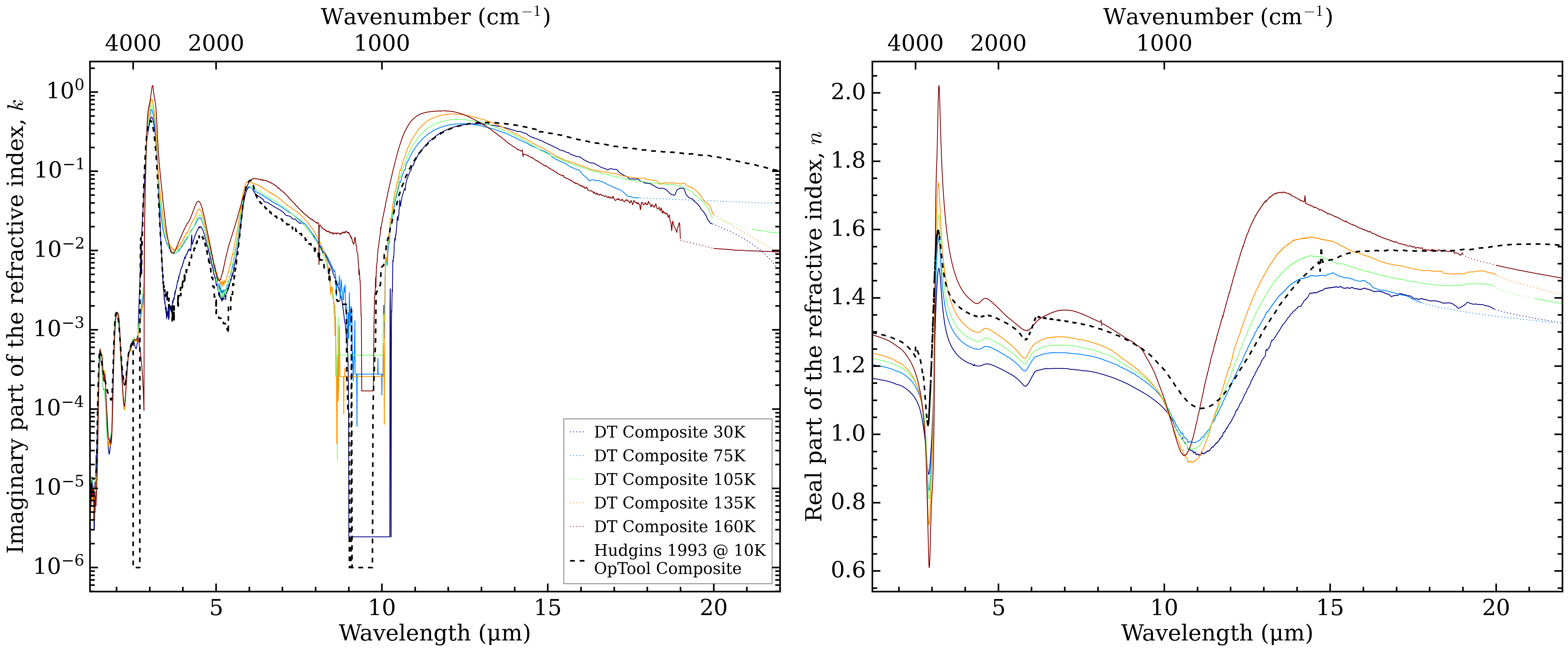}
      \caption{NIR - MIR zoom-in of 1.2 - 20\micron\ of the \textbf{\textit{DT}} OCCs at each temperature shown together with the \citet{hudgins_mid-_1993} at 10~K data. The dotted regions indicate where the $k$ value interpolation has been performed. 
              }
         \label{fig:DT_1.2_20}
\end{figure*}

The data we use to describe the 3\micron\ \ce{H2O} ice band in our \textbf{\textit{DT}} OCCs are the $k$ values from Leiden Ice Database for Astrochemistry (LIDA) \citep{rocha_lida_2022}. LIDA contains $k$ values for five fixed temperatures of 30~K, 75~K, 105~K, 135~K and 160~K. The 30~K, 75~K, 105~K and 135~K measurements were taken by \citet{oberg_effects_2007} and the 160~K data from \citet{gerakines_ultraviolet_1996} both covering 2.5 - 20\micron. For both of their experimental setups, they deposit water vapour onto a cesium iodide (CsI) substrate and took measurements at the fixed deposition temperature. The $k$ values were calculated from the experimentally measured absorption spectra by \citet{rocha_water_2024}. There is a small range of negative values from $\sim$8.5 - 10.3\micron, which is unphysical. The origin of the negative values is not explained in \citet{rocha_water_2024}, therefore, we replace them with a constant value of 10$^{-3}$ as the largest neighbouring, positive $k$ value across all temperatures. Using the constant $k$ value in place of the negative $k$ values has negligible impact on the calculated $n$ values whilst providing a reasonable estimate for physically possible $k$ values. The refractive indices are shown in Figure \ref{fig:DT_1.2_20}.

\subsubsection{2.64 - 2.82: Our interpolation}
\label{sec:DT_2.64_2.82}
A common cut off wavelength for every temperature in the \textbf{\textit{DT}} was required for consistency. 2.64\micron\ was determined to be an applicable wavelength to cut the \citet{mastrapa_optical_2009} $k$ values to include the \ce{H2O} overtones and exclude the beginning of the 3\micron\ absorption band profile. We do not use the \citet{rocha_water_2024} $k$ values directly from the 2.64\micron\ cut off because negative $k$ values are observed in this region for the 75~K, 105~K and 135~K measurements. Instead, we decided to interpolate the $k$ values from \citet{mastrapa_optical_2009} at 2.64\micron\ for all temperatures for consistency through all temperatures. We interpolate to 2.82\micron\ for all temperatures except 30~K, where we interpolate to 2.75\micron\ for the best profile match. We use log - log interpolation for all temperatures except for 160~K, where we use linear interpolation for the best profile match. 

\subsubsection{18 - 22\micron: Temperature-specific interpolation}

Log($\nu$) - log($k$) interpolation is performed to smoothly transition from the \citet{rocha_water_2024} data to the FIR experimental dataset for each temperature. Each interpolation is shown in Appendix \ref{sec:App_DT_NIRFIR}. For 30~K and 135~K, the interpolation is performed between 20 - 22\micron, 105~K between 20 - 21\micron, 75K is extended to 18 - 22\micron\ due to a mismatch in the \citet{rocha_water_2024} and \citet{smith_molecular_1994} and 160~K interpolates between 19-20\micron\ because the \citet{rocha_water_2024} is very noisy beyond 19\micron.

\subsubsection{20 - 74.5\micron: \citet{smith_molecular_1994} \& \citet{curtis_measurement_2005}}
\label{sec:DT_20_74.5}

We use two different experimental datasets in the FIR wavelength range for our \textbf{\textit{DT}} OCCs. The first is \citet{smith_molecular_1994}'s deposit and measure at a fixed temperatures of 30~K and 70~K, used in our 30~K and 75~K \textbf{\textit{DT}} OCCs respectively. In their experiments, they deposit \ce{H2O} vapour onto a low-density \ce{(C2H4)_n} substrate to build their ice and collect data from 20 - 110\micron. The second dataset is \citet{curtis_measurement_2005}'s deposit and measure at a fixed temperatures of 106~K, 136~K and 156~K used in our 105~K, 135~K and 160~K \textbf{\textit{DT}} OCCs respectively. They deposited \ce{H2O} vapour onto a silicon substrate to build their ices and collect data from 15 - 192\micron.

\begin{figure*}[t]
   \centering
   \includegraphics[width=0.85\textwidth]{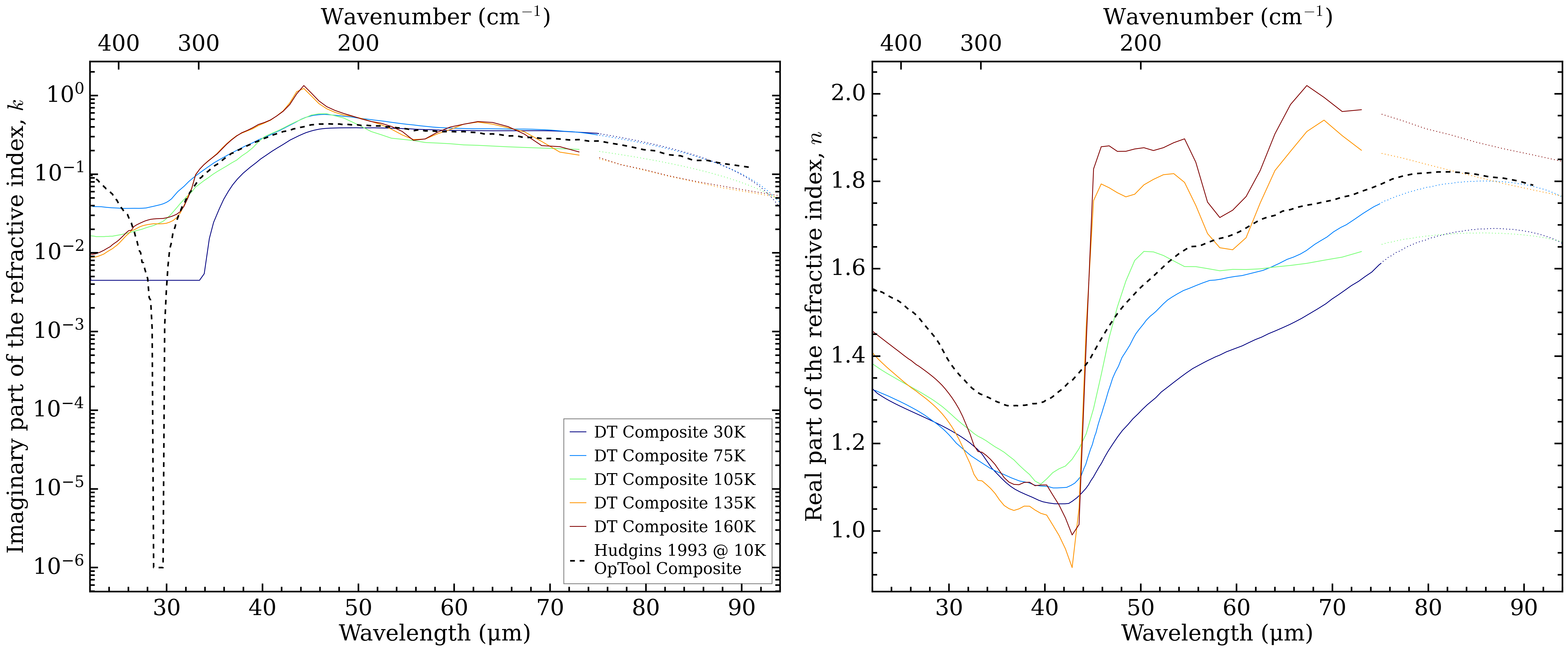}
      \caption{MIR - FIR zoom-in of 15 - 94\micron\ of the \textbf{\textit{DT}} OCCs at each temperature shown together with the \citet{hudgins_mid-_1993} at 10~K data. The dotted regions indicate where the $k$ value interpolation has been performed.}
         \label{fig:DT_15_94}
\end{figure*}

Figure \ref{fig:DT_15_94} shows the data for all temperatures, where clear temperature dependence is observed. However, there may be slight differences to true temperature dependent behaviour in this region due to the two different experimental setups. Both the \citet{smith_molecular_1994} and \citet{curtis_measurement_2005} datasets become noisy from around 100\micron. This motivated our decision to cut the experimental data at 94\micron\ and use the $k$ values of the temperature-dependent analytical model of \citet{reinert_absorption_2015}, as discussed in Section \ref{sec:94_10000}. 

\subsubsection{74.5 - 94\micron: Our interpolation}
\label{sec:DT_74.5_94}
For the 30~K, 75~K and 105~K, additional interpolation at the longer wavelengths is required. At 94\micron, the \citet{smith_molecular_1994} \textbf{\textit{DT}} data are at much higher values than the \citet{reinert_absorption_2015} model $k$ values. This may be a thermal history / ice phase effects between the two datasets. This discontinuity produces an unphysical sudden drop of the continuum of any spectra derived by RT modelling and therefore, must be addressed. It is noted that \citet{smith_molecular_1994} \textbf{\textit{DT}} data at 130~K measures higher $k$ values than an equivalent dataset, \citet{curtis_measurement_2005} \textbf{\textit{DT}} data at 136~K measurements, suggesting this is not a thermal history effect but an experimental issue. We deem that the \citet{smith_molecular_1994} $k$ values are overestimated in this region and perform interpolation between \citet{smith_molecular_1994} and \citet{reinert_absorption_2015} $k$ values. 

To best match the profile of the experimental data, linear interpolation between 74.5\micron\ for the \citet{smith_molecular_1994} 30~K and 75~K data and the start of the \citet{reinert_absorption_2015} $k$ values at 94\micron\ is chosen. Additionally, the \citet{curtis_measurement_2005} measurements show larger $k$ values than the \citet{reinert_absorption_2015} models between at 94\micron\ region. This difference may be a true reflection on the effect of thermal history on this wavelength region, due to the inconsistency between \citet{curtis_measurement_2005} (\textbf{\textit{DT}}) and \citet{reinert_absorption_2015} (\textbf{\textit{DC}}). To preserve the temperature dependence in the 94 - 10,000\micron\ region and to not have a sharp decrease in $k$ values at 94\micron, we performed linear interpolation between 74.5\micron\ in the 105~K \citet{curtis_measurement_2005} data to 94\micron, see figures in Appendix \ref{sec:App_DT_FIRsubmm}. There is excellent alignment between \citet{curtis_measurement_2005} 136~K and 156~K with the \citet{reinert_absorption_2015} models at the same temperatures and no interpolation is required.

\subsection{Deposited and cooled (\textbf{\textit{DC}}) measurement pure \ce{H2O} OCCs}
\label{sec:DC_exps}

\subsubsection{1.2 - 15\micron: \citet{mastrapa_optical_2009}}
\label{sec:DC_1.2_15}
We use experimental data of \citet{mastrapa_optical_2008,mastrapa_optical_2009} for the NIR region of our \textbf{\textit{DC}} OCCs. In their experiments, \ce{H2O} vapour was deposited at 150~K onto either a potassium bromide (KBr), zinc selenide (ZnSe) or cesium iodide (CsI) substrate, then cooled, with spectroscopic measurements from 1.1 - 15\micron\ taken at 10~K increments from 150~K down to 20~K. Figure \ref{fig:DC_0.1_1.2} shows the lack of temperature dependence in this region for the \textbf{\textit{DC}} OCCs. This highlights the need for experiments to test whether the $n_{632\textrm{nm}}$ value is temperature dependent for  cooled down \ce{H2O} ices, as it is for fixed temperature \ce{H2O} ice \citep{he_refractive_2022}.

\begin{figure*}[t]
   \centering
   \includegraphics[width=0.85\textwidth]{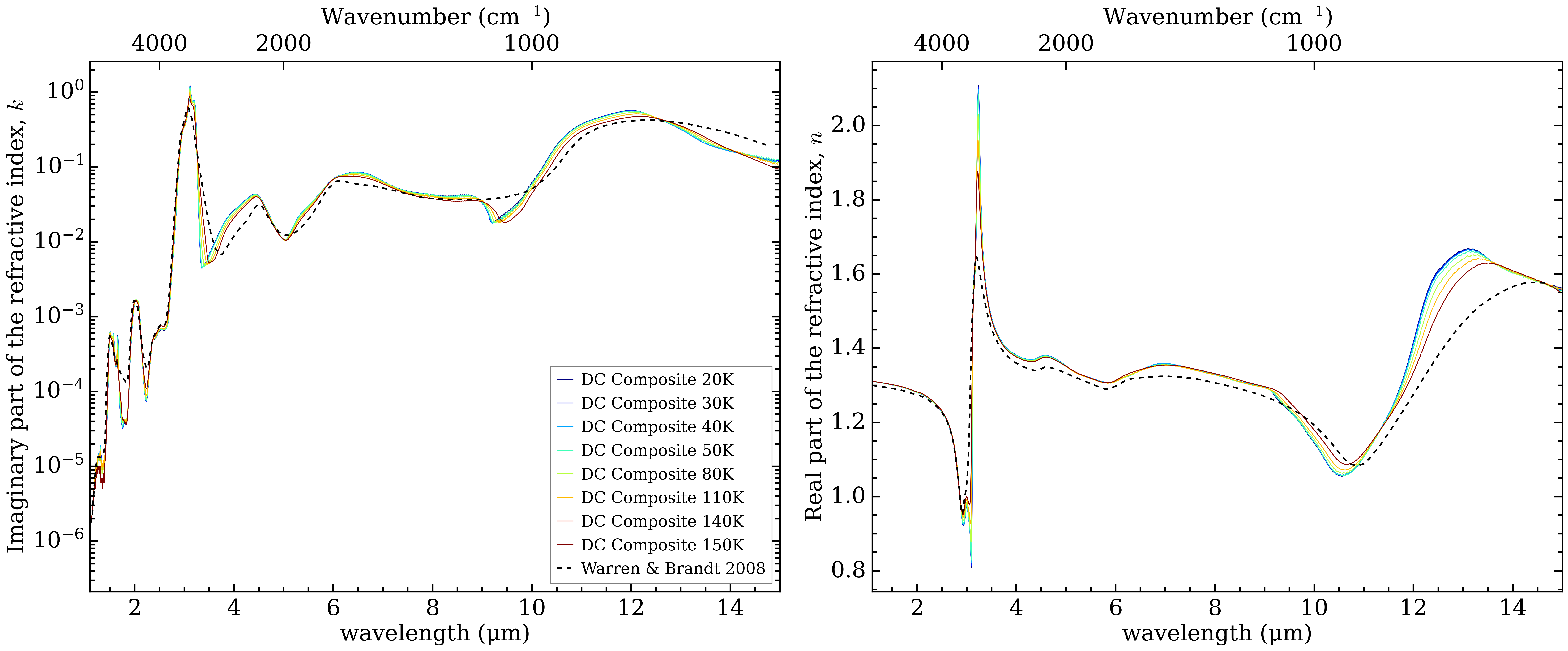}
      \caption{NIR - MIR zoom-in of 1.1 - 15\micron\ of the \textbf{\textit{DC}} OCCs at each temperature shown together with the \citet{warren_optical_2008} at 266~K data. Our OCCs use the deposited and cooled measurements of  \citet{mastrapa_optical_2009}.}
         \label{fig:DC_1.2_15}
\end{figure*}

\subsubsection{20 - 74.5\micron: \citet{smith_molecular_1994}}
\label{sec:DC_20_94}
\begin{figure*}[t]
   \centering
   \includegraphics[width=0.85\textwidth]{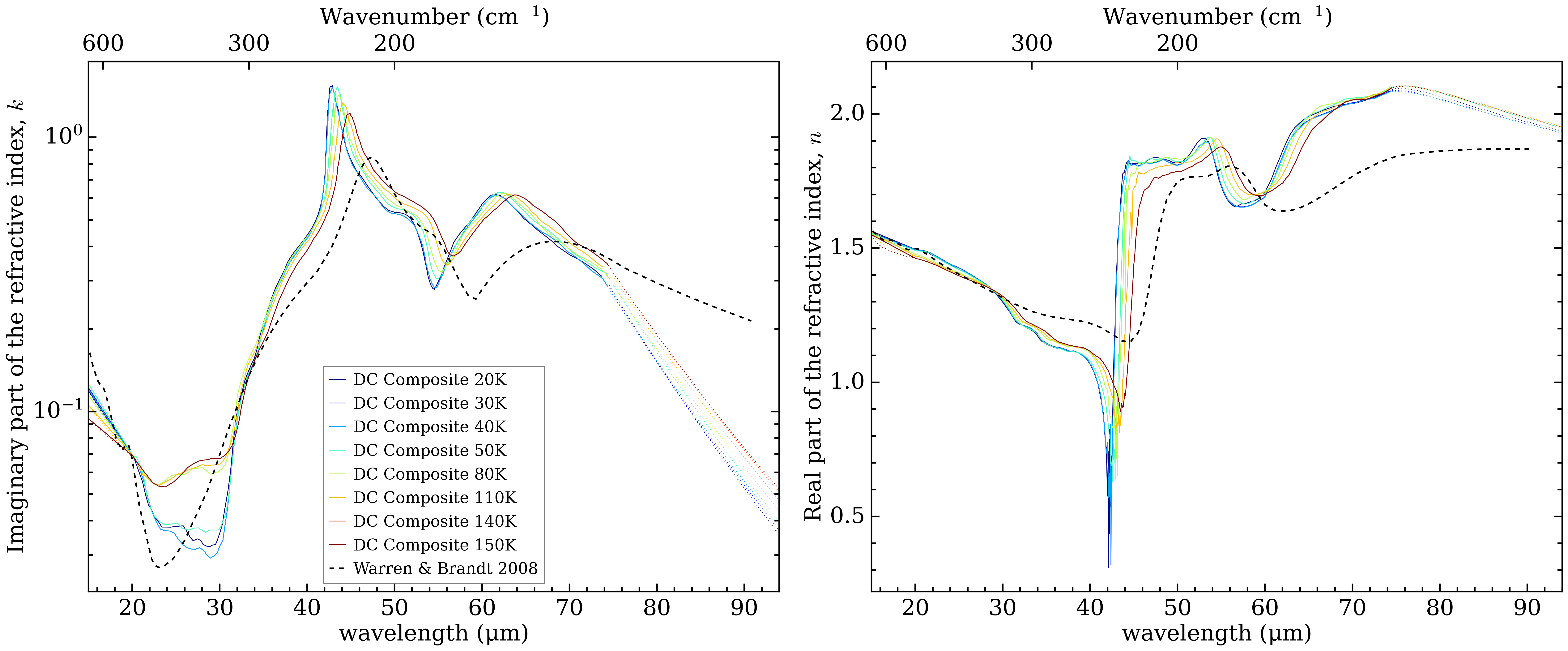}
      \caption{MIR - FIR zoom-in of 15 - 94\micron\ of the \textbf{\textit{DC}} OCCs at each temperature shown together with the \citet{warren_optical_2008} at 266~K data. All our OCCs use the deposited and cooled measurements of \citet{smith_molecular_1994}. The dotted regions indicate where the $k$ value interpolation has been performed.
              }
         \label{fig:DC_15_94}
\end{figure*}

In the FIR wavelength range, we use the \citet{smith_molecular_1994} cool down series of measurements, which covers 20 - 110\micron. In these experiments, \ce{H2O} vapour is deposited directly onto a low-density \ce{(C2H4)_n} substrate, preheated at 140~K and then is cooled with a series of measurements taken at 140, 130, 120, 110, 100, 90, 80, 70, 50, 30 and 10~K. This means that for our 20~K, 40~K and 150~K \textbf{\textit{DC}} OCCs, we do not have a direct temperature match between the NIR and FIR datasets. Thus, we choose to use the \citet{smith_molecular_1994} data that is 10~K lower than the NIR dataset temperature. A large discrepancy in $k$ values at 94\micron\ with \citet{smith_molecular_1994} \textbf{\textit{DC}} data showing much larger $k$ values than the \citet{reinert_absorption_2015} \textbf{\textit{DC}} models. This suggests this is not a thermal history issue, as mentioned in Section \ref{sec:DT_74.5_94}, but rather a experiment setup up issue.  Due to this, we perform an interpolation. Figure \ref{fig:DC_15_94} shows the experimental data up to 7.45\micron\ along with the interpolation from 74.5 - 94\micron\ for all \textbf{\textit{DC}} temperatures.

\subsubsection{15 - 20\micron: Our interpolation}
\label{sec:DC_15_20}
There is missing experimental data between 15 - 20\micron\ for the \textbf{\textit{DC}} thermal history. To overcome this, we employ the same technique as \citet{warren_optical_1984} to connect datasets and interpolate $k$ values between the two datasets in log($\nu$) - log($k$) space. The interpolation between the two datasets are shown in the figures of Appendix \ref{sec:App_DC_NIRFIR}. 

\subsubsection{74.5 - 94\micron: Our interpolation}
\label{sec:DC_74.5_94}
We perform an log($\nu$) - log($k$) interpolation between the \citet{smith_molecular_1994} data and  \citet{reinert_absorption_2015} models which retains the peak and spectral profile of the 63\micron\ emission feature observed in the \citet{smith_molecular_1994} data whilst also providing a smooth continuum through to 94\micron\ where the \citet{reinert_absorption_2015} $k$ values begin. We start the interpolation from the experimental data point nearest to 74.5\micron\ through to 94\micron. The interpolation between the two datasets are shown in the figures of Appendix \ref{sec:App_DC_FIRsubmm}.

\section{Construction of multi-temperature ice RT models}
\label{sec:ModelSetup}
Our thermal-history and temperature-dependent optical constant composite sets allow us to create the first series of RT disk models with self-consistent \ce{H2O} ice optical constants across the full required wavelength range. The HH 48 NE disk's edge-on geometry (inclination >70\degr) allows us to probe the \ce{H2O} ice using its 3\micron\ ice absorption and 45 and 63\micron\ emission features simultaneously as a background continuum for the absorption feature is provided by the central protostar and its scattered light. This combination of the updated disk model and new globally self-consistent OCCs provides a novel opportunity to:

\begin{itemize}
    \item Study the effects of ice thermal history on each of these feature's profiles in unison and begin to test the region of the disk contributing to each of these features
    \item Whether including crystalline ice only within a temperature-appropriate region ($>$125~K in Table \ref{tab:Complete_Region_Setup}) is enough to produce a crystalline profile without additional dynamical processes
    \item Model the effect of dynamical processes on ice spectral profiles through unique distributions of thermally processed and non-thermally processed material throughout a disk
\end{itemize} 

Upcoming FIR observatories will enable us to compare the impact of ice thermal history on both the NIR and FIR ice features and model different dynamical processes through unique ice distributions within RT models. This work provides an essential tool in identifying excellent candidates for FIR observatories based on previous observations. Such sample selection will be critical given the small number of targets that can be observed during these mission's lifetimes.
\\

To build this first series of self-consistent ice RT models, we use the radiative transfer code \texttt{RADMC-3D} \citep{dullemond_radmc-3d_2012} with its full anisotropic scattering capabilities enabled. Our fiducial model is a disk continuum setup from \citet{sturm_jwstmiri_2024}, where they used JWST NIRSpec IFU, MIRI MRS, HST scattered light images, ALMA and archival spectral energy distribution (SED) data to derive the disk dust density structure and stellar properties. Their dust distribution employs the two dust grain size populations (small: <12.63\micron, large: >12.63\micron\ - <3000\micron) with the same dust grain material properties used in \citet{sturm_jwstmiri_2024}. 

\begin{figure}[t]
   \centering
   \includegraphics[width=\linewidth]{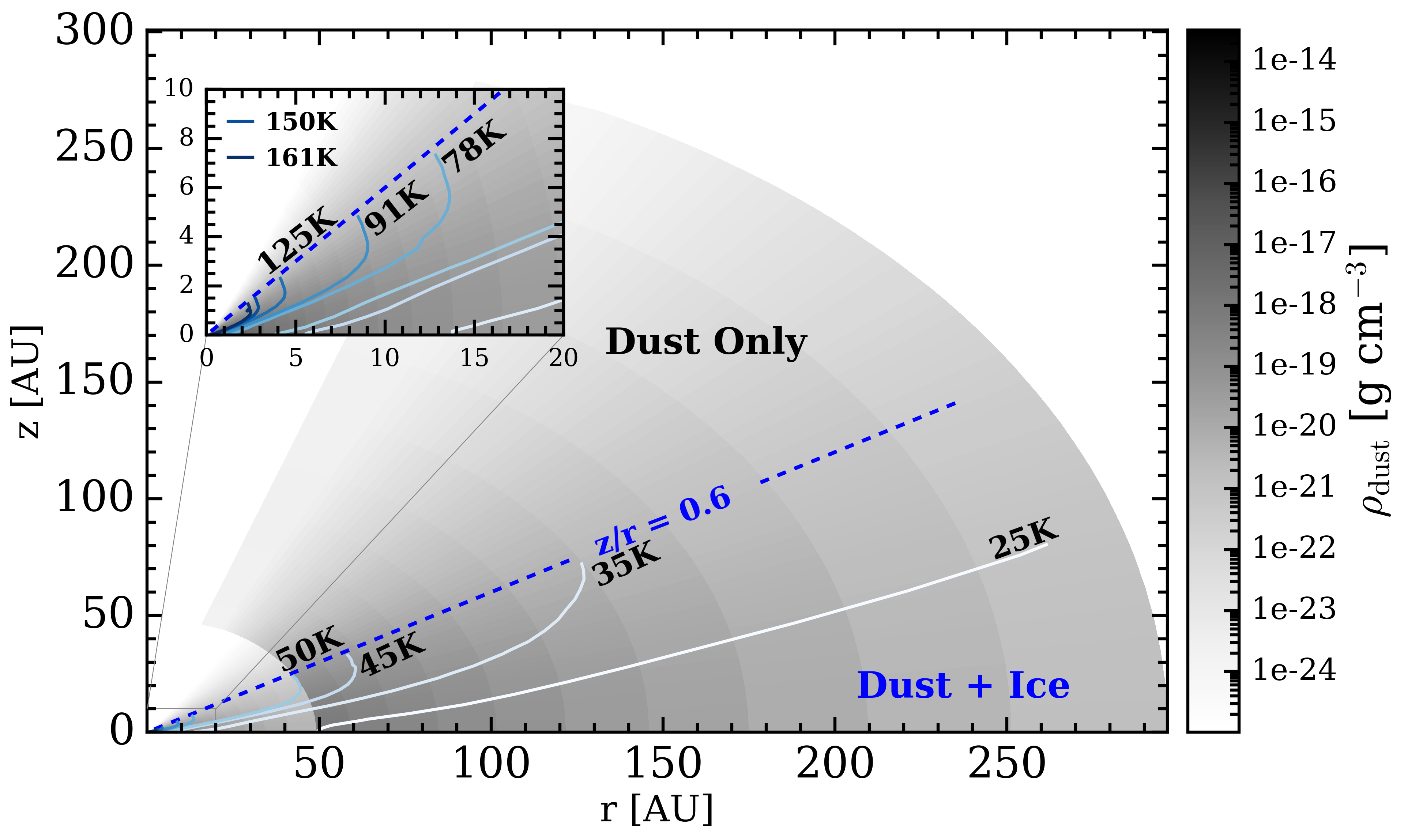}
      \caption{Temperature regions within our disk models. The bold black outline shows the full extent of the dust within our disk models. The z/r = 0.6 dashed line shows the limit set for ices.}
         \label{fig:temp_regs}
   \end{figure}

We add an ice mixture with constant ice abundance fractions w.r.t \ce{H2O} ice derived by \citet{sturm_jwstmiri_2024} for HH 48 NE onto our dust grains. Our ice distribution throughout the disk uses a mix of approaches of \citet{sturm_edge-protoplanetary_2023} and \citet{sturm_jwstmiri_2024}. The ices have a limited extent through the disk due to photodesorption. This limit is set to include ices on the dust in regions of z/r $\leq$ 0.6. The distribution of our ice species follows the approach of \citet{sturm_edge-protoplanetary_2023} where the ices were distributed according to their pure ice desorption temperatures, as opposed to by their binding energies as done in \citet{sturm_jwstmiri_2024}. This is to ensure that our newly derived temperature-dependent optical constants are used in the appropriate regions of the disk. The ice mixture includes a 3:1 mix of \ce{CO2} and CO (15\% and 5\% w.r.t \ce{H2O} respectively), \ce{NH3} (1.7\%) and \ce{CH4} ($<$12\%) ices in addition to \ce{H2O}. We employ the optical constants provided by \texttt{OpTool} for \ce{NH3} and \ce{CH4} and measurements by \citet{bergner_jwst_2024} for the \ce{CO2} and CO mix. This ensures that any continuum effects from the additional ices are modeled, allowing a true comparison with the 3\micron\ spectral profile of \citet{sturm_jwstmiri_2024}'s model. Using this ice distribution approach and ice mixture, we are able to split the disk in to eleven regions when using the \textbf{\textit{DC}} set and seven regions using the \textbf{\textit{DT}} set. These regions and their according ice species are listed in Table \ref{tab:Complete_Region_Setup} and the full eleven regions are shown in Figure \ref{fig:temp_regs} via their dust temperature contours. 

\begin{table*}[t]
    \centering
    \caption{Ice mixture and temperatures per region using \textbf{\textit{DC}} and \textbf{\textit{DT}} \ce{H2O} OCCs.}
    \label{tab:Complete_Region_Setup}
    \setlength{\tabcolsep}{3pt} 
    \renewcommand{\arraystretch}{0.95}
    \begin{tabular}{l l l l l}
    \hline\hline
    Region & MDC Ice & Opacity & MDT Ice  & Opacity \\
    $T$ range (K) & Species & Reg. \# & Species & Reg. \# \\
    \hline
    $>$161      & Bare dust grains & 10 & Bare dust grains & 6 \\
    161 - 147   & \ce{H2O}[150] & 9  & \ce{H2O}[160] & 5 \\
    147 - 125   & \ce{H2O}[140] & 8  & \ce{H2O}[135] & 4 \\
    125 - 91    & \ce{H2O}[110] & 7  & \ce{H2O}[105] & 3 \\
    91 - 78     & \ce{H2O}[80], NH$_3$ & 6  & \ce{H2O}[75], NH$_3$   & 2 \\
    78 - 70     & \makecell[l]{\ce{H2O}[80], NH$_3$, CO$_2$:CO 3:1} & 5 & \ce{H2O}[75], NH$_3$, CO$_2$:CO 3:1 & \multirow{2}{*}{\bigg\}1} \\
    70 - 50     & \ce{H2O}[50], NH$_3$, CO$_2$:CO 3:1 & 4 & \ce{H2O}[75], NH$_3$, CO$_2$:CO 3:1 \\
    50 - 45     & \makecell[l]{\ce{H2O}[50], NH$_3$, CO$_2$:CO 3:1, CH$_4$} & 3 & \ce{H2O}[30], NH$_3$, CO$_2$:CO 3:1, CH$_4$ & \multirow{4}{*}{\(\left\}\vphantom{\begin{array}{c}
        X\\X\\X\\X
        \end{array}}\right.\)0} \\
    45 - 35 & \ce{H2O}[40], NH$_3$, CO$_2$:CO 3:1, CH$_4$ & 2 & \ce{H2O}[30], NH$_3$, CO$_2$:CO 3:1, CH$_4$ &  \\
    35 - 25 & \ce{H2O}[30], NH$_3$, CO$_2$:CO 3:1, CH$_4$ & 1 & \ce{H2O}[30], NH$_3$, CO$_2$:CO 3:1, CH$_4$ &  \\
    25 - 0  & \ce{H2O}[20], NH$_3$, CO$_2$:CO 3:1, CH$_4$ & 0 & \ce{H2O}[30], NH$_3$, CO$_2$:CO 3:1, CH$_4$ &  \\
    \hline
    \end{tabular}
     \tablefoot{The temperatures have been best matched to each region and in each phase but an exact match is not always possible. The temperature of the \ce{H2O} ice in each region is given in square brackets. The opacity region number columns correlate to those in Fig. \ref{fig:opacities}.}
    
\end{table*}

To test the impact of the self-consistent models, we perform an initial study of the original \ce{H2O} ice OCCs against our self-consistent models. Table \ref{tab:RT_Models} outlines the 4 models we built for this test. Model M0 and M0A use the standard single ice temperature and phase OCCs from \texttt{OpTool}; model M0 uses \citet{warren_optical_2008} \ce{H2O} \cry\ at 266~K and model M0A uses \citet{hudgins_mid-_1993} \ce{H2O} \am\ at 10~K data for the NIR and FIR wavelength regions of interest to this study. M0 is the model fitted in \citet{sturm_jwstmiri_2024}. We know that disks have radial temperature gradients and this widely used approach is not a realistic representation of ice throughout the disk. The other two models include our self-consistent \ce{H2O} ice optical constants. Model MDT uses our \textbf{\textit{DT}} composite series, where crystalline ice is only present in the region of the disk that is 125~K and above. This is determined by the temperature structure calculations from \texttt{RADMC-3D}. Figure \ref{fig:temp_regs} shows that this region is deeply embedded in the disk and extends out to 3~AU based on the HH 48 NE system parameters. The rest of the ice in the disk is in the amorphous phase. This the view of a disk without any additional dynamical processes transport or forming crystalline ice in unexpected outer regions. The final model, MDC, uses our \textbf{\textit{DC}} composite series, meaning the entire disk is populated by crystalline ice which has either been transported from the hot inner 3~AU region or has been crystallised in-situ through a thermal processing event, such as a global episodic accretion outburst or local planetesimal collisions. This ice will have subsequently cooled, retaining its crystalline structure. Theoretically, each of these crystallisation routes would produce different distributions of crystalline ices, but modelling each phenomena individually is outside the scope of this paper. This model could test the reset vs. inheritance scenario for sufficiently young disks; however, the age of class II disks is typically beyond the 0.5 Myr timescale on which re-amorphisation of the ices by external high energy radiation or particles occurs \citep{grundy_distributions_2006,cook_near-infrared_2007, mcclure_detections_2015}. Additionally, bulk crystallisation of the entire outer disk through infall and spreading has yet to be demonstrated by chemical models incorporating dynamics \citep{visser_chemical_2011,drozdovskaya_cometary_2016}.

\begin{table}[h]
    \centering
    \caption{Model names and their according \ce{H2O} ice OCCs.}
    \begin{tabular}{l l c}
        \hline
        \hline
        \makecell{Model \\ Name}  & Optical Constants & \makecell{\ce{H2O} Ice \\ Properties} \\
        \hline
        M0 & \citet{warren_optical_2008} & 266~K \cry \\
        M0A & \citet{hudgins_mid-_1993} & 10~K \am \\
        MDC & Our \textbf{\textit{DC}} series & \makecell{Temp-dep. \\ cooled \cry}\\
        MDT & Our \textbf{\textit{DT}} series & \makecell{Temp-dep. fixed \\  temperature ices} \\
        \hline
    \end{tabular}
    \label{tab:RT_Models}
\end{table}

\begin{figure*}[t]
    \centering
    \includegraphics[width=0.85\textwidth]{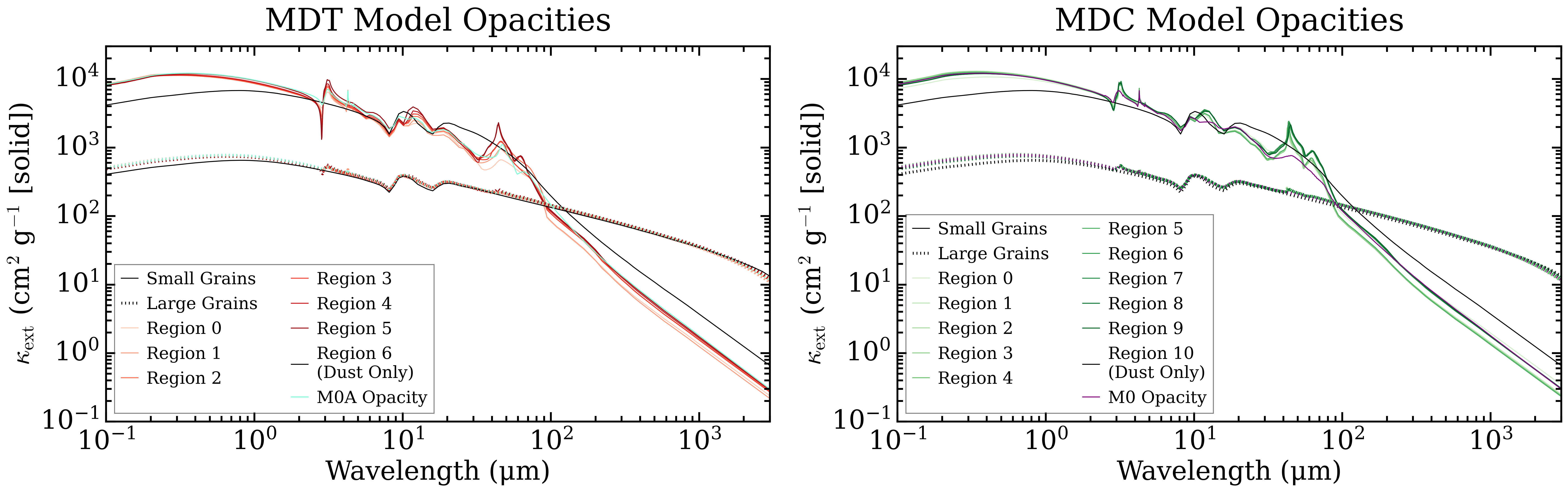}
    \caption{The opacities of the ice mixtures (see Table \ref{tab:Complete_Region_Setup}) from cool down and deposition temperature optical constants. Left: Regional ice mixture opacities using the eight (\textbf{\textit{DC}}) \ce{H2O} OCCs. Right: Regional ice mixture opacities using the five (\textbf{\textit{DT}}) \ce{H2O} OCCs.}
    \label{fig:opacities}
\end{figure*}

In Figure \ref{fig:opacities}, we present the small and large grain population opacities for each of the eleven possible ice mixtures including the cool down \ce{H2O} optical constants and the seven including the deposition temperature \ce{H2O} optical constants we use within our models. The changes in spectral profile for the 3, 45 and 63\micron\ \ce{H2O} features between each measurement technique, and thus ice phase, are evident from the opacities. We see the small population opacities dominate at the 3, 45 and 63\micron\ \ce{H2O} features wavelengths. We kept a constant grain size distribution for the dust only and dust and ice opacities, meaning the dust only opacities have larger extinction then the dust and ice opacities in some wavelength regimes. 

\section{Results}
\label{sec:Results}

We discuss the impact that self-consistently applying our new temperature-dependent \ce{H2O} OCCs to our expanded temperature-zoned RT models makes on the spectral profiles of the strongest water ice features, at 3, 45 and 63\micron. We compare the spectral profiles to the same edge-on disk model of HH 48 NE using the single temperature OCCs widely-used in the literature \citep[e.g.][]{warren_optical_2008, hudgins_mid-_1993}. 
We compare our output 3\micron\ profiles to the JWST NIRSpec IFU observations of HH 48 NE, as seen in the left panel of Figure \ref{fig:OpTvsTDOCs} for each of the four models and make predictions of the spectral profiles of the  45/63\micron\ features for future FIR space missions (right panel) .
Our results indicate that these features can effectively probe the thermal history of water ice. 

\subsection{3\micron\ ice absorption feature}
\label{sec:NIRres}

\subsubsection{Comparisons with HH 48 NE observations}
\label{sec:Model_HHobs}
\begin{figure*}[t]
   \centering
   \includegraphics[width=0.85\textwidth]{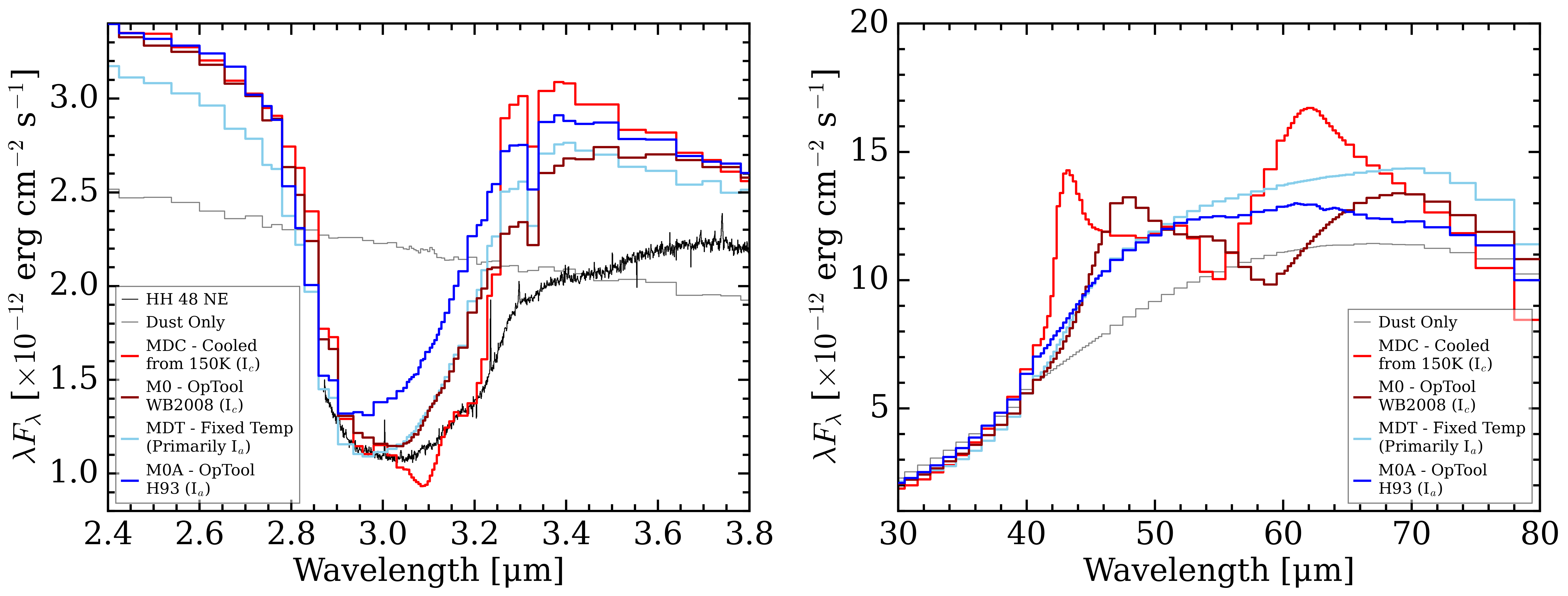}
      \caption{The 3\micron\ \ce{H2O} ice absorption feature and 45/63\micron\ \ce{H2O} ice emission feature spectral profiles from \texttt{RADMC-3D} models using the single temperature OCCs from \citet{warren_optical_2008} (\ce{H2O} \cry\ - M0) and \ce{H2O} \am\ based on \citet{hudgins_mid-_1993} (M0A) and our temperature-dependent \textbf{\textit{DT}} and \textbf{\textit{DC}} OCCs. These models use the HH 48 NE disk input parameters as determined by \citet{sturm_jwstmiri_2024,sturm_edge-protoplanetary_2023-1}, therefore a profile comparison with the JWST NIRSpec IFU spectra of HH 48 NE \citep{sturm_jwst_2023} is shown.}
    \label{fig:OpTvsTDOCs}
\end{figure*}

The M0 model fits the depth of the 3\micron\ ice feature but not necessarily the profile. In \citet{sturm_jwst_2023}, they suggest that the observed 3.2\micron\ feature in the JWST NIRSpec IFU data, shown in Figure \ref{fig:OpTvsTDOCs}, could be an indication of crystalline ice in the disk. However, if the ices in HH 48 NE are crystalline, then using the \citet{warren_optical_2008} crystalline ice at 266~K, as in the M0 model, is unlikely to match the profile, because the maximum sublimation temperature for water ice is 160~K under typical protoplanetary disk conditions. 

Our MDC model, with the temperature-dependent optical constants and their self-consistent zoning throughout the RT model provides a better diagnostic into whether the 3.2\micron\ feature arises in crystalline ice. We see that although the MDC model matches the depth of the 3.2\micron\ feature given the fitted disk parameters, it also produces an extremely strong 3.085\micron\ peaked feature which is completely absent from the observations. Since the bottom of the 3\micron\ feature is clearly not saturated and past observations of edge-on disks \citep{terada_discovery_2012,terada_adaptive_2012} have shown that the depth of the 3.1\micron\ peak is always stronger than the 3.2\micron, then the 3.2 \micron\ feature in HH 48 NE is better explained by some other solid state absorption, rather than by crystalline water ice.

We see that the M0A model of the fully amorphous disk and the MDT model of the disk with warmed-up ices distributed according to the intrinsic temperature structure (which are mostly amorphous disk beyond 3~AU) both produce a very similar red wing shape to the M0 model. The lack of 3.1 and 3.2\micron\ features in the \citet{warren_optical_2008} crystalline ice composite demonstrates it is unsuitable to match the known crystalline 3\micron\ disk ice profiles from the literature. The blue wing of the observations does not show as sharp of a slope as the M0 or MDC models, matching the amorphous M0A and MDT models more closely. This suggests that the \ce{H2O} ice in HH 48 NE is mostly amorphous. 

The depth of the 3\micron\ feature in model M0A does not match the observations well because \citet{sturm_jwstmiri_2024} derived abundances using the \citet{warren_optical_2008} optical constants to match the profile. Using the amorphous optical constants from \citet{hudgins_mid-_1993} would have required a higher abundance to fit the depth of the ice band in the observations. In the MDT model, the continuum underlying the 3 \micron\ feature is somewhat lower in flux. This offset is due to the temperature-dependence of n$_{632nm}$ discussed in Section \ref{sec:OCs}. For the same disk parameters, this would alter the SED fit and final ice abundances that would be derived. These results highlight the essential dependence of \ce{H2O} ice abundances on accurate optical constant measurements and their self-consistent application in RT models. 

\subsubsection{Locating the absorbing ices}
\label{sec:Det_3_Reg}

We would like to use RT models to locate the absorbing ices within disks. Previously this has been accomplished by constructing a contribution function of the ices to the observed feature, which determines the region in which the ice originates within the model, and therefore its temperature. This has been done analytically for certain RT models (e.g. Equation 3 of \citet{mcclure_detections_2015}). Though this is difficult to do with Monte Carlo codes like \texttt{RADMC-3D} because backtracing scattered photons is not practical. However, in principle we should be able to use the temperature information encoded in the observed spectral profiles to provide constraints on the ices' location.

The MDT model, where crystalline ice exists only in regions hot enough for amorphous ice to crystallise (>125~K), displays no crystalline absorption. This demonstrates that for HH 48 NE, which has typical classical T Tauri stellar and disk parameters, the only way to observe crystalline ice would be for it to be present over a large area of the outer disk. For crystalline ice to exist there, additional dynamical processes must have transported it there from the inner disk or generated it locally through heating events. This finding motivates further study of similar edge-on classical T Tauri disks with crystalline ice 3\micron\ profiles, such as HK Tau B \citep{terada_discovery_2012} and d216-0939 \citep{terada_discovery_2012,potapov_simple_2025} to determine the origin of their crystalline ice profiles. This is not necessarily the case for Herbig Ae/Be or and F/G type intermediate mass T Tauri disks, as their hotter central stars may allow ice to crystallise at more distant radii than in classical T Tauri disks and be observed without additional dynamical ice crystallisation mechanisms.

Although the peak wavelength shift from 3.1\micron\ at 150~K to 3.12\micron\ at 20~K in the \citet{mastrapa_optical_2009} seems like a promising metric to locate the crystalline ices beyond the 125~K isotherm from the observations, our MDC model shows that scattering induced by the disk geometry causes the peak wavelength to fall outside of this range, at 3.085\micron. This means we cannot use the peak wavelength of the 3\micron\ feature as a metric to infer a location of the absorbing ices. The importance of scattering to the 3\micron\ profile was also demonstrated empirically by \citet{martinien_role_2025}, where they showed that the peak wavelength of the 3\micron\ feature shifts to shorter wavelengths at higher disk inclinations. HH 48 NE is at an inclination of 83\degr, an angle at which we expect significant scattering to occur, as demonstrated by \citet{sturm_jwstmiri_2024}.

\subsection{45 and 63\micron\ ice emission features}
\label{sec:FIRres}

There are no observations of HH 48 NE in the FIR range to compare our models with, but we discuss below the application of these model to potential future FIR ice observations. 

The right panel of Figure \ref{fig:OpTvsTDOCs} shows our four models. The M0 model exhibits broad emission features peaking at $\sim$47 and $\sim$70\micron, consistent with \citet{sturm_jwstmiri_2024}. In contrast, the temperature-dependent MDC model produces features at 43.2 and 62.2\micron, matching the expected peak wavelengths from \citet{smith_molecular_1994} for thermally processed crystalline \ce{H2O} ice that has cooled to $<$80~K. We can use these peak wavelengths to constrain the ice's emission region. Figure \ref{fig:opacities} shows that the 45 and 63\micron\ features are dominated by the small-grain population, with $a_{\max}$=12.59\micron, placing the contributing grains in the Rayleigh regime at these wavelengths. In this regime, scattering is negligible and therefore the peak wavelengths shift only with temperature, not with grain size. Instead, Figure 3 of \citet{mcclure_probing_2012} demonstrated that while the 43\micron\ feature is very sensitive to the power of the grain size distribution, both features become flatter and develop red wings in response to increasing maximum grain size, such that the feature must trace maximum sizes smaller than 30\micron. Figure 6 of \citet{kamp_diagnostic_2018} confirms this finding for the \citet{smith_molecular_1994} optical constants as well. Therefore, the agreement between the MDC model peak wavelengths and the \citet{smith_molecular_1994} cooled crystalline ice measurements implies that the emitting ice in our model share the same thermally processed and subsequently cooled ($<$80~K) crystalline structure as in the laboratory experiments. This means ices in the disk atmosphere within $\sim$13 AU of the central star do not contribute to the FIR features in our disk models (see Figure \ref{fig:temp_regs}). Locating the exact radial origin of the strongest emission from our RT models would still require some kind of contribution function as in \citet{mcclure_detections_2015} or \citet{sturm_jwstmiri_2024}.

The M0A model shows the clear lack of emission features in the FIR region. This is as expected as these features are specifically a result of crystalline \ce{H2O} ice. The MDT model completely emulates the shape of the continuum of the M0A in the FIR wavelength range with a very slight increase in flux.

\section{Discussion}
\label{sec:Discussion}

Our globally self-consistent \ce{H2O} ice OCCs will now enable modellers to retrieve the full insight across each \ce{H2O} ice feature. Being able to predict spectral profiles of all the \ce{H2O} ice features using our OCCs and temperature-specific opacity zoning model setup will have far-reaching implications for RT disk modelling and NIR/FIR observational planning.
\\

Firstly, our study demonstrates that constraints on the crystalline ice distribution can be found even using a single feature. We show that strong crystalline ice signatures in either the NIR and FIR feature profiles, for HH 48 NE's disk parameters, must arise from ice residing in cool, outer disk regions, far from crystalline ice's in-situ forming region. These results strongly support the findings of \citet{mcclure_detections_2015, min_abundance_2016} and \cite{kamp_diagnostic_2018} using the 45 and 63\micron\ features alone, where the FIR crystalline ices were all found to originate in the outer regions of disks as a result of dynamical processing occurring within the disks. That we see the same effect for the NIR ices in our disk model implies that dynamical processes may also be the cause of ice crystallisation in classical (K0 and later spectral type) T Tauri disks, more generally. 

Combining our model setup and OCCs should enable us to vary the ice's thermal history through the disk, allowing us to emulate the effects of different local dynamical processes such as viscous spreading or planetesimal collisions. We explore this in Smith et al. (in prep.), where we will vary the ice crystallinity throughout the disk model through abundance ratios throughout the entire disk but also through varying the spatial distribution of crystalline \ce{H2O} ice throughout the disk. Such models may identify spectral profile indicators associated to each dynamical process and, therefore, greatly enhance the interpretation of edge-on disk observations. 

Unfortunately, very few objects have both NIR and FIR ice observations. Comparative samples will not become available until a FIR observatory becomes operational. These models and OCCs linking the 3\micron\ and 45/63\micron\ features will aid in the interpretation of archival NIR and FIR observations, new JWST NIR observations and future FIR missions, such as the Planetary Origins and Evolution Multispectral Monochromator (POEMM), PRobe far-Infrared Mission for Astrophysics (PRIMA) and the Far-Infrared Great Observatory (FIGO). The ability to accurately select disks that will show crystalline ice features for future observations in the NIR or FIR will require zoned RT models, such as ours, to reveal what information each feature provides observers.

\section{Conclusions}
\label{sec:Concs}
In this study, we build the first set of self-consistent \ce{H2O} ice OCCs across the full wavelength range required for RT modelling. We have done so for ices with two different thermal histories: deposited-and-cooled (\textbf{\textit{DC}}) and deposited-and-measured at a fixed temperature (\textbf{\textit{DT}}). By doing so we have improved the derivation of the $n$ values for \ce{H2O} ice using data across a broad wavelength range, while identifying wavelengths which require new experimental data. These updates enable the derivation of accurate \ce{H2O} ice opacities across many different physical and chemical environments. Our OCCs have been uploaded to a Zenodo repository, DOI: \href{https://zenodo.org/records/20313567}{10.5281/zenodo.2031356}. 

We have also updated the \texttt{RADMC-3D} RT model of the edge-on classical T Tauri disk,  HH 48 NE \citep{sturm_jwstmiri_2024}, to include temperature zoning appropriate to our new OCCs. With these new RT models, we have demonstrated self-consistent ice analysis across the full RT model wavelength, improving compared to previous local to FIR self-consistent modelling efforts from \citet{mcclure_probing_2012,mcclure_detections_2015,min_abundance_2016} and \citet{kamp_diagnostic_2018}. Our models clearly show that observations of the NIR and FIR ice profiles can pinpoint the origin of crystalline ice features in the cold outer regions of classical T Tauri disks, below their nominal crystallisation temperature. These results indicate that classical T Tauri disks with observed crystalline \ce{H2O} profiles must be host to additional dynamical processes such as viscous spreading, accretion event heating or planetesimal collisions. The combination of our new OCCs and updated RT models enable us to begin testing how these different dynamical processes can be traced via the profiles of the 3, 45 and 63\micron\ ice features.

This work provides the basis for self-consistent \ce{H2O} ice modelling within RT models allowing modellers to begin to trace the thermal and structural evolution of \ce{H2O} ice from molecular clouds through to planetary systems. Essentially, these OCCs provide the basis for accurate planning of analysis of observations of complementary datasets across multiple instruments.

\begin{acknowledgements}
Z.L.S., M.K.M., I.K. and S.C. acknowledge support from grant ASTRO.JWST.001 (Dutch Astrochemistry in the era of JWST - DAN-III) from the Dutch Research Council (NWO). M.K.M acknowledges financial support from the NWO Talent Programme Vidi (grant number 233.122) from NWO.
\end{acknowledgements}

\bibliographystyle{aa}
\bibliography{refs}

\onecolumn
\begin{appendix}
\section{Interpolation Regions}
\label{sec:AppA}

\subsection{\textbf{\textit{DT}}}
\label{sec:AppA_DT}

\subsubsection{2.64 - 2.82\micron\ interpolation}

\begin{figure*}[!htbp]
   \centering
   \includegraphics[width=\linewidth]{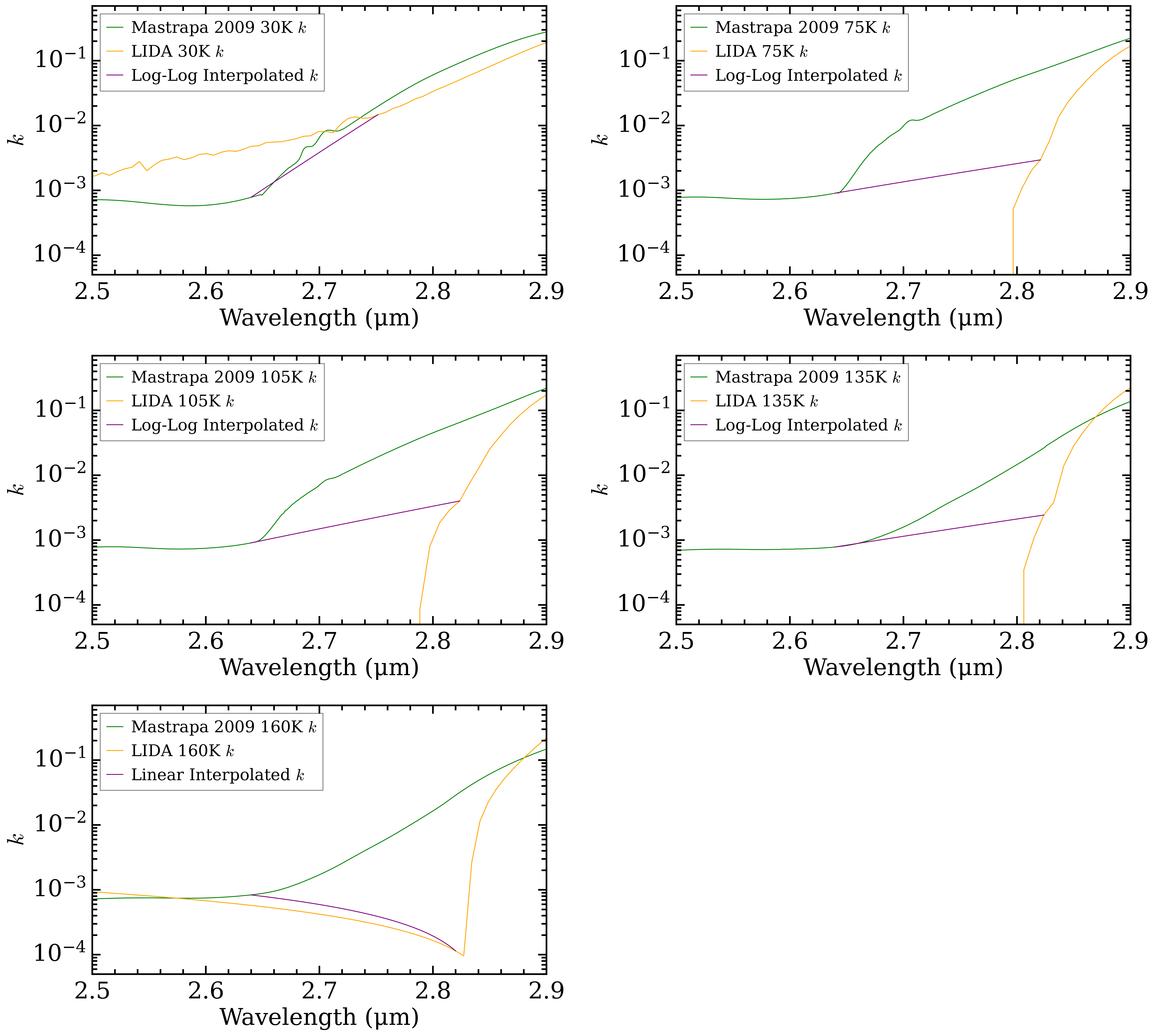}
   \caption{$k$ value interpolation between the \citet{mastrapa_optical_2009} and \citet{rocha_water_2024} NIR data of the \textbf{\textit{DT}} OCC series. Note the 30~K interpolation only reaches 2.75\micron\ instead of 2.82\micron. Also note that the 160~K composite uses linear interpolation rather than log-log interpolation.}
   \label{fig:DT_interp_vis_NIR}
\end{figure*}

\newpage
\subsubsection{NIR - FIR interpolation}
\label{sec:App_DT_NIRFIR}

\begin{figure*}[!htbp]
   \centering
   \includegraphics[width=\linewidth]{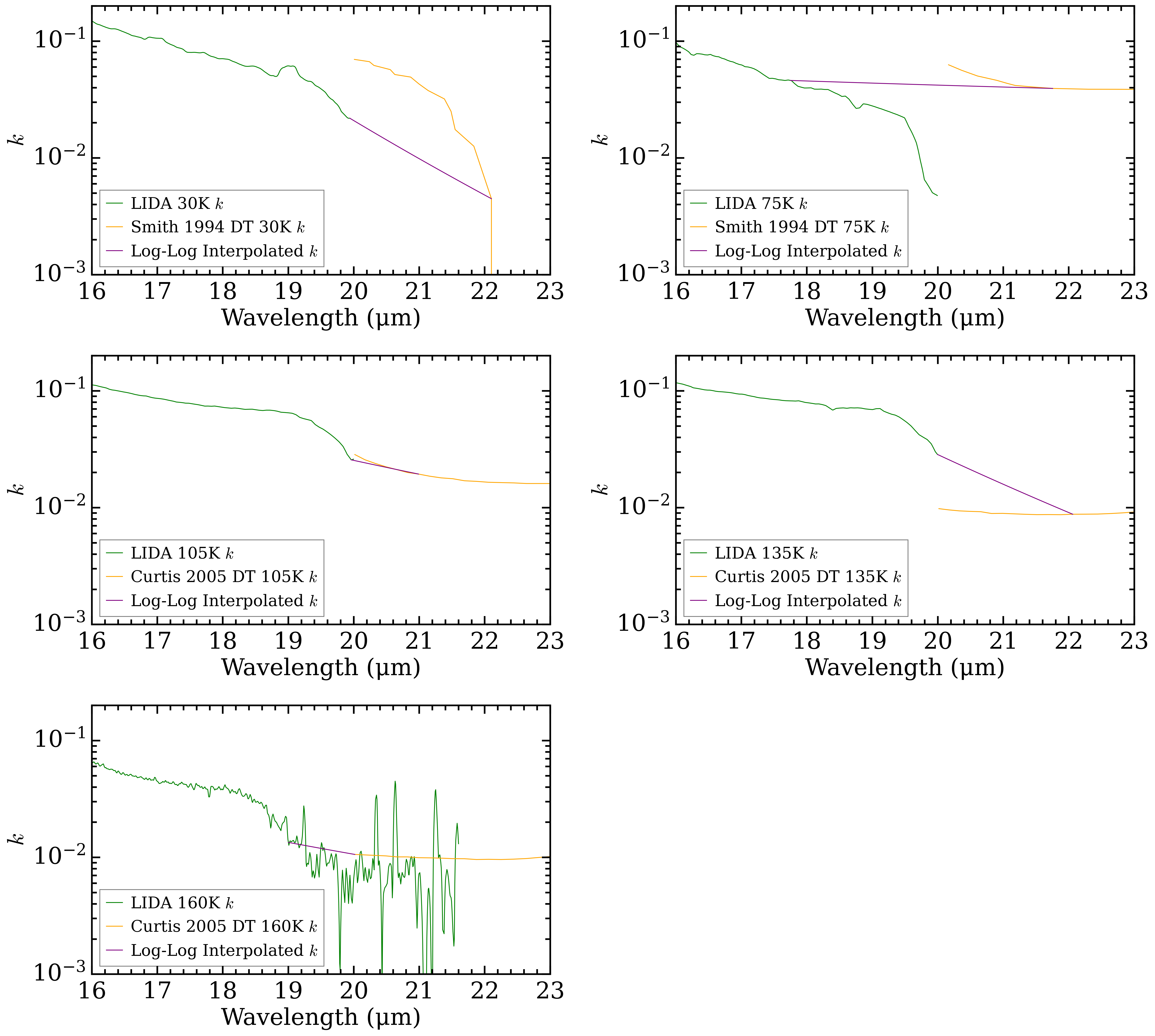}
   \caption{$k$ value interpolation between the end of the \citet{rocha_water_2024} NIR data and the start of the \citet{smith_molecular_1994} data of the \textbf{\textit{DT}} composites.}
   \label{fig:DT_interp_NIR_FIR}
\end{figure*}

\newpage
\subsubsection{74.5 - 94\micron\ interpolation}
\label{sec:App_DT_FIRsubmm}

\begin{figure*}[!htbp]
   \centering
   \includegraphics[width=\linewidth, height=0.85\textheight, keepaspectratio]{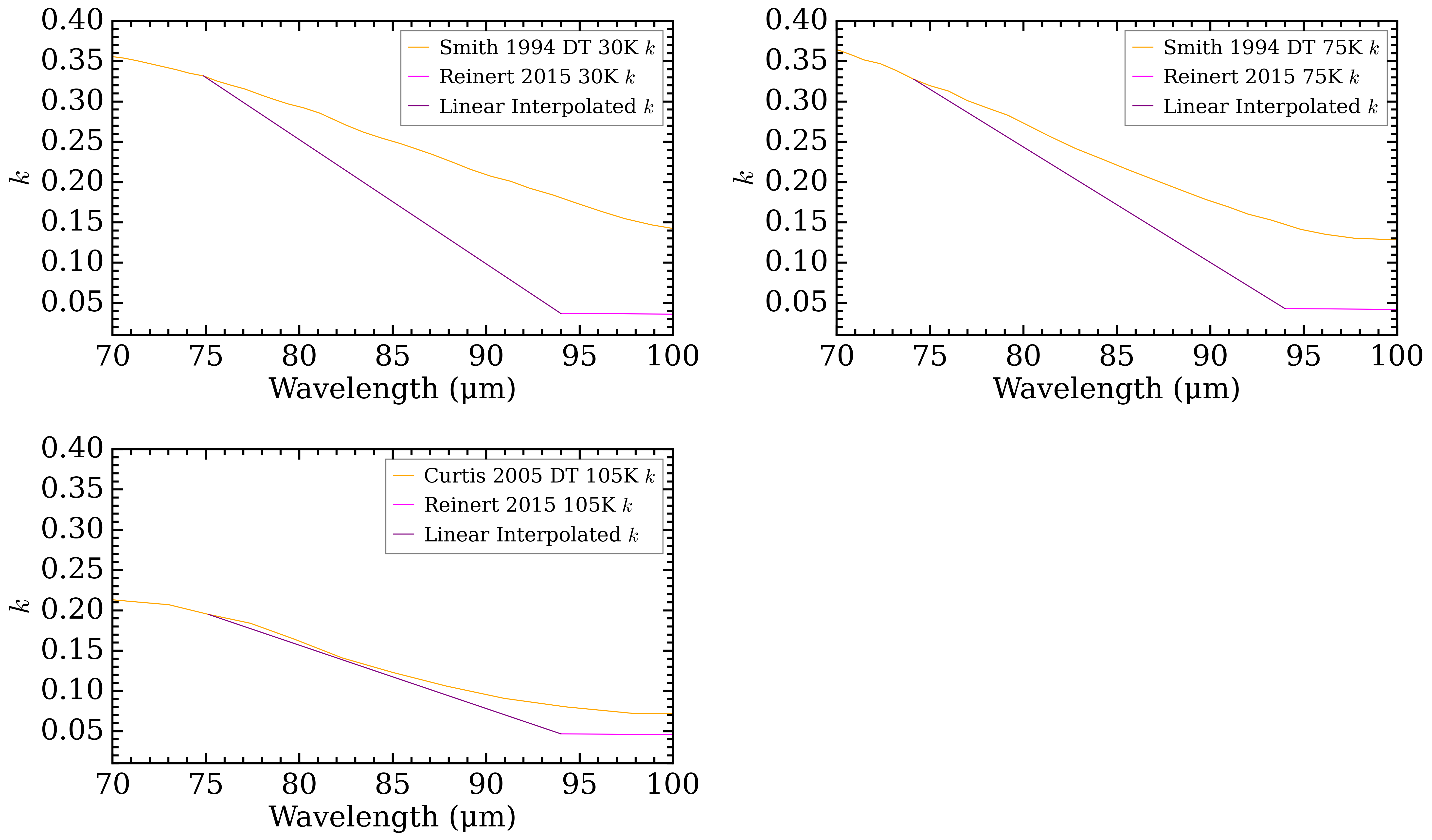}
   \caption{$k$ value interpolation between the end of the \citet{smith_molecular_1994} FIR data and the start of the \citet{reinert_absorption_2015} model of the \textbf{\textit{DT}} 30~K, 75~K and 105~K composites. The 135~K and 160~K required no interpolation.}
   \label{fig:DT_interp_FIR_submm}
\end{figure*}

\newpage
\subsection{\textbf{\textit{DC}}}
\label{sec:AppA_DC}

\subsubsection{15 - 20\micron\ interpolation}
\label{sec:App_DC_NIRFIR}

\begin{figure}[!htbp]
   \centering
   \includegraphics[width=\linewidth, height=0.85\textheight, keepaspectratio]{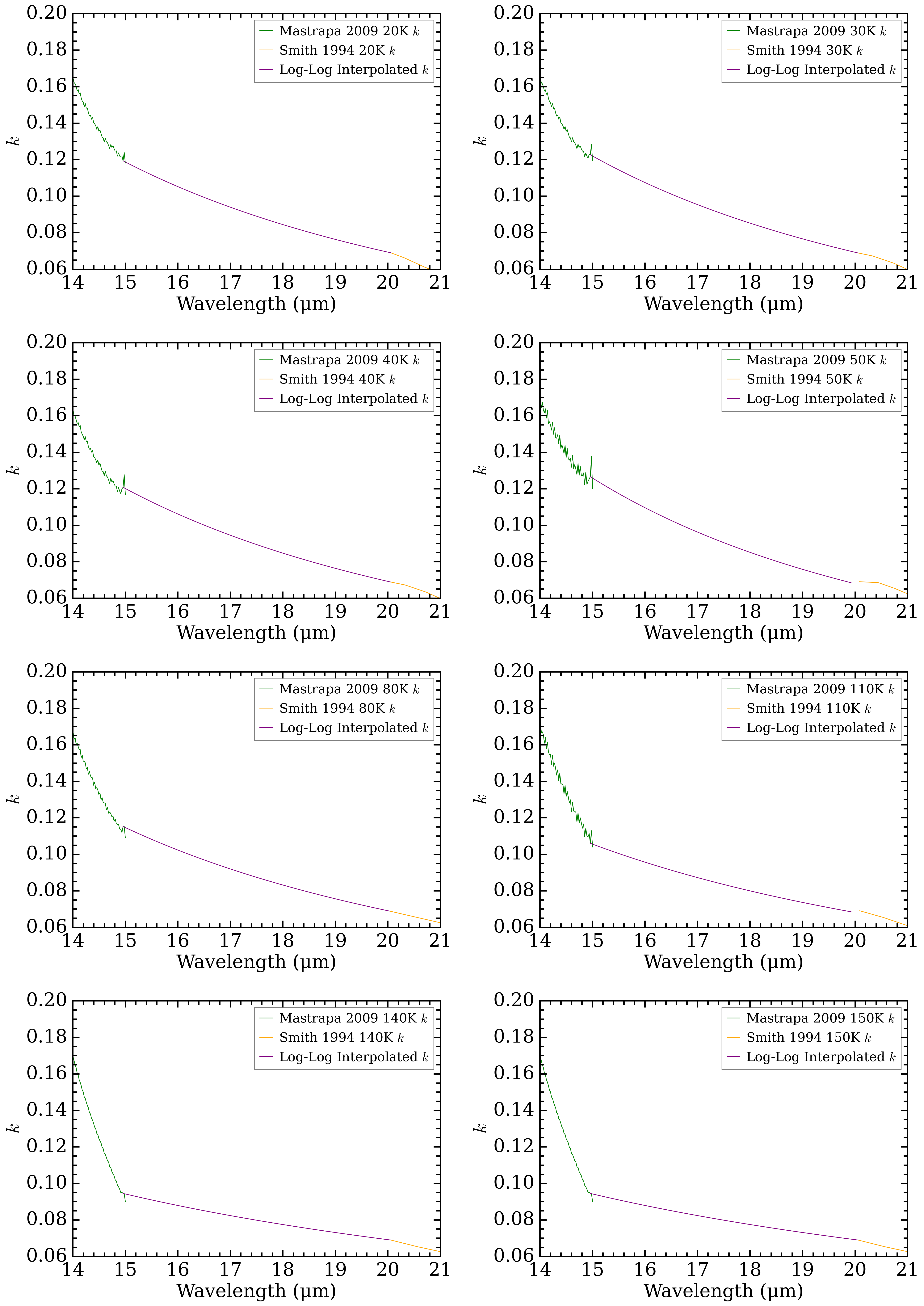}
   \caption{$k$ value interpolation between the end of the \citet{mastrapa_optical_2009} NIR data and the start of the \citet{smith_molecular_1994} FIR data of the \textbf{\textit{DC}} composite.}
   \label{fig:DC_interp_NIR_FIR}
\end{figure}

\newpage
\subsubsection{74.5 - 94\micron\ interpolation}
\label{sec:App_DC_FIRsubmm}

\begin{figure}[!htbp]
   \centering
   \includegraphics[width=\linewidth, height=0.85\textheight, keepaspectratio]{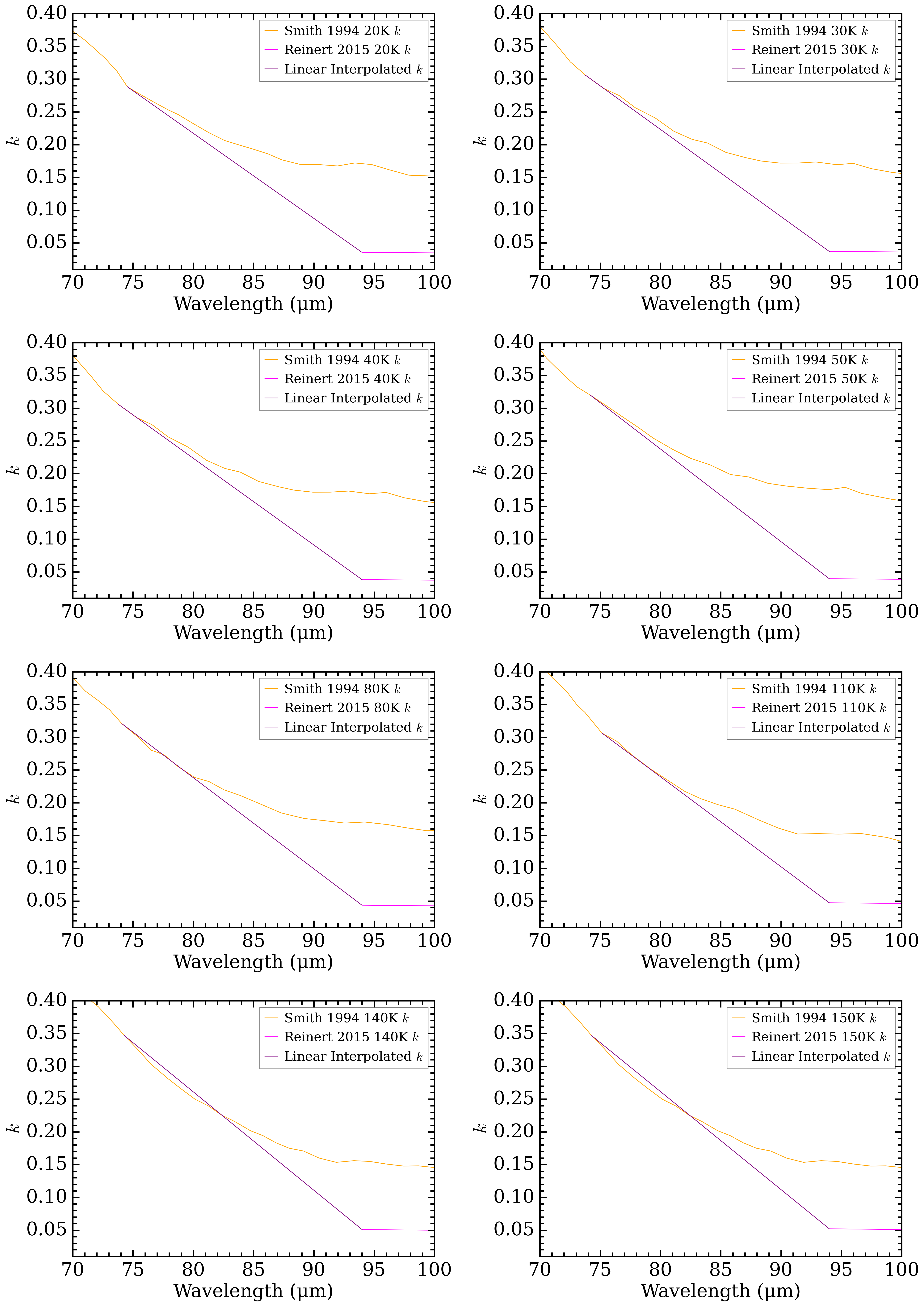}
   \caption{$k$ value interpolation between the end of the \citet{smith_molecular_1994} FIR data and the start of the \citet{reinert_absorption_2015} model for all \textbf{\textit{DC}} composites.}
   \label{fig:DC_interp_FIR}
\end{figure}

\end{appendix}
\end{document}